\documentclass[12pt]{iopart}

\pdfoutput=1

\usepackage[toc,page]{appendix}
\usepackage{graphics}
\usepackage{graphicx}
\usepackage{epsfig}
\usepackage{epstopdf}

\usepackage[table]{xcolor}
\usepackage{url}
\usepackage{bm}
\usepackage{mathrsfs}

\usepackage{silence}
\usepackage{hyperref} 
\hypersetup{
    colorlinks=true,
    linkcolor=blue,
    citecolor=blue,
    filecolor=magenta,
    urlcolor=cyan,
}
\usepackage{color} 
\usepackage{soul}
\usepackage{tikz}
\usetikzlibrary{calc}

\usepackage{pgfplots}
\pgfplotsset{width=5.7cm,compat=1.9}
\pgfplotsset{yticklabel style={text width=2.5em,align=right}}
\pgfplotsset{every axis/.append style={thick}}
\tikzset{every picture/.style={font=\fontsize{9}{10.8}\selectfont}}

\usepackage{xspace}
\newcommand{\BHaH}{{\texttt{BlackHoles@Home}}\xspace}
\newcommand{\bhah}{{\texttt{BH@H}}\xspace}
\newcommand{\ET}{{\texttt{Einstein Toolkit}}\xspace}
\newcommand{\et}{{\texttt{ET}}\xspace}
\newcommand{\nrpy}{{\texttt{NRPy}}\xspace}
\newcommand{\nrpycuda}{{\texttt{NRPy-CUDA}}\xspace}
\newcommand{\superb}{{\texttt{superB}}\xspace} 
\newcommand{\superbnrpy}{{\texttt{superB/NRPy}}\xspace} 
\newcommand{\openmp}{{\texttt{OpenMP}}\xspace} 
\newcommand{\groovy}{{\texttt{GRoovy}}\xspace} 
\newcommand{\charmpp}{{\texttt{Charm++}}\xspace}
\newcommand{\twopunctures}{{\texttt{TwoPunctures}}\xspace}
\newcommand{\ckio}{{\texttt{Ckio}}\xspace}
\newcommand{\projections}{{\texttt{Projections}}\xspace}

\newcommand\be{\begin{equation}}
\newcommand\ee{\end{equation}}

\newcommand\bse{\begin{subequations}}
\newcommand\ese{\end{subequations}}
\newcommand\bea{\begin{eqnarray}}
\newcommand\eea{\end{eqnarray}}

\newcommand{\pd}{\partial}
\newcommand{\Lie}{\mathcal{L}} 

\newcommand\ringring[1]{%
  {
   \mathop{\kern0pt #1}\limits^{
     \vbox to-1.85ex{
       \kern-2ex 
       \hbox to 0pt{\hss\normalfont\kern.1em \r{}\kern-.45em \r{}\hss}%
       \vss 
     }
   }
  }
}
\begin{document}

\title[\superbnrpy Numerical Relativity Framework for 3G Gravitational Wave Science]{\superbnrpy: Scalable, Task-Based Numerical Relativity for 3G Gravitational Wave Science}

\author{Nishita Jadoo$^*$}
\address{Department of Physics, University of Idaho, Moscow, ID 83843, USA}
\ead{njadoo@uidaho.edu}

\author{Terrence Pierre Jacques}
\address{Department of Physics and Astronomy, West Virginia University, Morgantown, WV 26506, USA}
\address{Center for Gravitational Waves and Cosmology, West Virginia University, Chestnut Ridge Research Building, Morgantown, WV 26505, USA}
\address{Department of Physics, University of Idaho, Moscow, ID 83843, USA}
\ead{tp0052@mix.wvu.edu}

\author{Zachariah B.~Etienne}
\address{Department of Physics, University of Idaho, Moscow, ID 83843, USA}
\address{Department of Physics and Astronomy, West Virginia University, Morgantown, WV 26506, USA}
\address{Center for Gravitational Waves and Cosmology, West Virginia University, Chestnut Ridge Research Building, Morgantown, WV 26505, USA}
\ead{zetienne@uidaho.edu}

\begin{abstract}
Modern gravitational-wave science demands increasingly accurate and computationally intensive numerical relativity (NR) simulations. The Python-based, open-source \nrpy framework generates optimized C/C++ code for NR, including the complete NR code \BHaH (\bhah), which leverages curvilinear coordinates well-suited to many astrophysical scenarios. Historically, \bhah was limited to single-node \openmp CPU parallelism. To address this, we introduce \superb, an open-source extension to \nrpy that enables automatic generation of scalable, task-based, distributed-memory \charmpp code from existing \bhah modules. The generated code partitions the structured grids used by \nrpy/\bhah, managing communication between them. Its correctness is validated through bit-identical results with the standard \openmp version on a single node and via a head-on binary black hole simulation in cylindrical-like coordinates, accurately reproducing quasi-normal modes (up to $\ell=8$). The \superbnrpy-generated code demonstrates excellent strong scaling, achieving an $\approx 45$x speedup on 64 nodes (7168 cores) compared to the original single-node \openmp code for a large 3D vacuum test. This scalable infrastructure benefits demanding simulations and lays the groundwork for future multi-patch grid support, targeting long inspirals, extreme parameter studies, and rapid follow-ups. This infrastructure readily integrates with other \nrpy/\bhah-based projects, enabling performant scaling for the general relativistic hydrodynamics code \groovy, and facilitating future coupling with GPU acceleration via the \nrpycuda project.
\end{abstract}

\maketitle

\footnotetext[1]{*Corresponding author: njadoo@uidaho.edu}

\section{Introduction}
\label{sec:introduction}

Numerical relativity (NR) serves as an essential tool for modeling strongly gravitating systems like binary black hole (BBH) and binary neutron star (BNS) coalescences. Its importance has grown significantly with the era of gravitational-wave (GW) astronomy, with many dozens of direct detections by the LIGO and Virgo detectors to date~\cite{LIGOScientific:2018mvr,LIGOScientific:2020ibl,LIGOScientific:2021usb,KAGRA:2021vkt}. These simulations, which solve the full Einstein equations numerically, yield vital predictions for GW signals, facilitating the extraction of astrophysical and fundamental physics insights from observations~\cite{Nitz:2021uxj,Baiotti:2016qnr,Shibata:2019wef}. In addition, NR waveforms are crucial for calibrating faster approximate waveform models~\cite{Pan:2009wj,Ramos-Buades:2023ehm,Khan:2015jqa,Field5,Varma:2019csw,Albanesi:2025txj} used extensively in GW data analysis pipelines.

Ongoing detector upgrades and the prospect of next-generation ground-based~\cite{Punturo2010,Reitze:2019dyk,Evans2021} and space-based observatories~\cite{TianQin2015,AmaroSeoane2017,Babak2021} promise a dramatic increase in GW detections and observable phenomena. This progress demands higher accuracy from NR waveforms~\cite{Purrer2020,Ferguson2021,Jan2024} and requires more efficient and scalable NR codes in order to produce long-duration GW signals, such as those from high-mass-ratio binaries~\cite{AmaroSeoane2018}. Developing next-generation NR codes optimized for high-performance computing (HPC) clusters is therefore paramount. This growing demand for accurate, scalable simulations motivates the development of new NR code generation frameworks capable of targeting distributed-memory HPC systems---the focus of this work.

A comprehensive list of scalable NR codes under active development can be found in Table 1 of Ref.~\cite{Foucart2022}. These codes can be classified according to their discretization schemes and grid structures. A significant group employs finite-difference methods with patch-based adaptive mesh refinement (AMR), including those utilizing the \texttt{CarpetX} \cite{Schnetter2004} framework in the \ET(\et) ~\cite{Loffler2012, EinsteinToolkit2024_11} such as \texttt{Llama} \cite{Pollney2011}, \texttt{McLachlan} \cite{Brown2009_2}, and \texttt{LEAN} \cite{Sperhake2007}, as well as \texttt{BAM} \cite{Brugmann2008}, \texttt{AMSS-NCKU} \cite{Cao2008}, and \texttt{GRChombo} \cite{Clough2015}. Other finite-differencing codes, like \texttt{Dendro-GR} \cite{Fernando2019,Fernando2022} and \texttt{GR-Athena++} \cite{Daszuta2021,Daszuta2024, Cook2025}, employ a block-based oct-tree AMR. Alternative discretization schemes include the pseudo-spectral techniques used in \texttt{SpEC} \cite{Szilagyi2009} and \texttt{bamps} \cite{Hilditch2016, Bugner2015}, and the hybrid finite-difference and discontinuous Galerkin methods implemented in \texttt{SpECTRE} \cite{Kidder2016, Spectre, Lovelace2025} and \texttt{Nmesh} \cite{Tichy2023}. Modern development trends include the adoption of task-based parallelism for dynamic load balancing and performance portability, as exemplified by codes like \texttt{SpECTRE} and \texttt{AthenaK} \cite{Hengrui2024}---an approach also adopted in this work---and automatic code generation via symbolic algebra, an approach facilitated by frameworks such as \texttt{Kranc} \cite{Husa2006}, \texttt{Simflowny} \cite{Palenzuela2021} and \nrpy~\cite{Ruchlin2018}.

Achieving stable, long-term simulations of inspiraling and merging BBHs represented major breakthroughs in NR~\cite{Pretorius:2005gq,Campanelli:2005dd,Baker:2005vv}. These successes relied heavily on robust evolution systems, such as the generalized harmonic formulation~\cite{GarfinkleGeneralizedHarmonicpre,Pretorius:2004jg}, as well as the BSSN formulation~\cite{Shibata1995,Baumgarte1998} and its covariant extension~\cite{Brown2009}, which are now widely adopted for their stability properties. Crucially, the development of the ``moving puncture'' technique~\cite{Campanelli:2005dd,Baker:2005vv}---which employs the 1+log slicing and Gamma-driver shift gauge condition to manage BH singularities and allow them to move across computational grids without requiring excision---proved essential for enabling routine BBH simulations.

Although NR codes have historically relied upon Cartesian grids with AMR, more recent efforts have explored curvilinear coordinates (e.g., spherical-like or cylindrical-like), which offer significant advantages for certain problems. Many astrophysical systems exhibit approximate symmetries, making curvilinear coordinates a natural choice. They can concentrate computational resources near regions of interest (such as BHs or neutron stars), simplify the implementation of outer boundary conditions for isolated systems, and potentially reduce artificial reflections found in AMR schemes~\cite{Zlochower2012,Etienne2014, Black:2025orbit}. However, adopting curvilinear coordinates introduces challenges, including the treatment of coordinate singularities and the complexity of tensor basis transformations~\cite{Baumgarte2013,Ruchlin2018}.

\nrpy~\cite{Ruchlin2018,nrpy_web} is an open-source, Python-based code generation framework designed to address such complexities. It builds on \texttt{sympy}~\cite{sympy} and its code generation capabilities to automatically convert high-level tensorial expressions in Einstein notation into highly optimized C/C++ code. This approach forms a foundation for a modular, maintainable, and high-performance NR infrastructure.

A key output of this framework is \BHaH~(\bhah)~\cite{Ruchlin2018,Etienne2024}, which should be understood not as a static program but as a complete, stand-alone NR code entirely generated by \nrpy. The generated \bhah code includes a full infrastructure for simulations---including modules for initial data, timestepping, single-patch numerical gridding, checkpointing, and diagnostics---all managed by a simple Makefile build system also constructed by \nrpy. The framework's power extends beyond stand-alone codes; \nrpy can also generate complete modules (``thorns'') for the established \et community framework, supporting both its legacy \texttt{Carpet} and modern \texttt{CarpetX} AMR drivers. All open-source \nrpy code, including the functionality to generate the \bhah examples discussed in this paper, is publicly available\footnote{See the GitHub repository at \url{https://github.com/nrpy/nrpy}. The `README.md' file provides instructions for generating numerous example codes.}. \nrpy-generated codes robustly handle coordinate singularities by treating the singular parts of tensor components analytically, so that only the regular (nonsingular) parts are ever interpolated or finite-differenced~\cite{Baumgarte2013,Montero2012,Ruchlin2018}.

Despite these capabilities, the application of \nrpy-generated codes to the most computationally demanding problems, which require HPC clusters, has been significantly limited. These codes, including \bhah, have largely relied upon shared-memory parallelism via \openmp. This restriction confined simulations to single compute nodes, preventing exploration of larger or more complex physical scenarios. Addressing this critical need for scalability is the primary motivation for our current work.

To overcome this limitation, we introduce \superb, an open-source extension to the \nrpy framework specifically designed to enable distributed-memory parallelism. \superb automatically generates scalable, task-based parallelized C++ code utilizing the \charmpp parallel programming system~\cite{kale1993charmpp,charmpp}. The resulting integrated framework, \superbnrpy, leverages the same underlying physics modules and infrastructure defined within \nrpy for \bhah, but extends them for distributed execution. \superbnrpy partitions the structured grids inherent to \bhah's design into parallel tasks (chares) managed by \charmpp. It automatically handles the necessary ghost zone communication between tasks, including the complex communication patterns required for both standard `outer' boundaries and the `inner' boundaries arising from coordinate singularities~\cite{Ruchlin2018}, using point-to-point messaging.

Our approach with \superbnrpy can be most closely compared to other HPC-oriented efforts in the NR community employing singular curvilinear coordinates. Notably, the closed-source \texttt{SphericalNR} framework~\cite{Mewes:2018szi,Mewes:2020vic,Ji:2023tok}, developed within the \et , performs dynamical-spacetime general relativistic magnetohydrodynamic (GRMHD) evolutions in spherical coordinates. Similar to \nrpy/\bhah, \texttt{SphericalNR} uses a reference-metric formalism~\cite{Brown2009,Montero2012} and tensor rescaling to manage coordinate singularities, evolving the BSSN or fCCZ4~\cite{Alic:2011gg} equations. It adapts the Cartesian GRMHD code \texttt{GRHydro}~\cite{Mosta:2013gwu} and made use of an earlier version of \nrpy~\cite{Ruchlin2018,senr_web} to generate specific source terms and spacetime kernels within the \et's existing MPI-parallelized Cactus/Carpet~\cite{Schnetter:2003rb,CarpetCode:web} infrastructure.

In contrast, our open-source \superbnrpy provides distributed-memory capabilities specifically for the \nrpy code generation framework. \superbnrpy supports a significantly broader range of curvilinear coordinate systems (including spherical-like, cylindrical-like, prolate-spheroidal-like, and Cartesian-like) and integrates with the general relativistic hydrodynamics (GRHD) code \groovy~\cite{Jacques:2024pxh}. Crucially, \superbnrpy automates the generation of \charmpp parallelized \bhah evolution codes directly from high-level symbolic inputs, offering significant advantages in flexibility, extensibility, and accessibility compared to integrating pre-existing code into larger frameworks.

We validate \superbnrpy by demonstrating numerical equivalence (bit-identical results on a single node) with the standard \openmp version of \bhah. To showcase its capabilities for demanding simulations, we simulate a head-on collision of two equal-mass, non-spinning BHs starting from a separation of $10\,M$ in cylindrical-like coordinates. We extract gravitational waveforms using the Weyl scalar $\psi_4$ and decompose the result into spin-weighted spherical harmonics, finding excellent agreement with quasi-normal mode (QNM) predictions up to $\ell=8$ during the ringdown phase. We further demonstrate compatibility with recent algorithmic improvements like the Slow Start Lapse (SSL) technique~\cite{Etienne2024}, which reduces constraint violations and is implemented directly within the shared \nrpy/\bhah framework.

Strong scaling tests confirm that \superbnrpy achieves excellent parallel performance. While the original \openmp code generated by \nrpy is more efficient on a single 16-core desktop node for a small 3D vacuum evolution test, the \charmpp-based version demonstrates a speedup of approximately 45x on 64 nodes (7168 cores) of a high-core-count cluster compared to the original single-node \openmp code for a large 3D vacuum evolution test. This scalability significantly expands the scope of NR applications accessible to the \nrpy ecosystem on modern HPC systems.

The HPC-capable parallel infrastructure developed in \superbnrpy establishes a foundation for numerous high-impact applications, including long-duration inspirals, extreme parameter studies, and rapid GW event follow-ups. It directly benefits other \nrpy-generated projects, such as the \groovy code, and paves the way for future integration with node-level GPU acceleration via the \nrpycuda project~\cite{Tootle:2025ikk}. This scalable \superb foundation will also be extended to support code generation for the multi-patch, multi-coordinate grid structures central to the \bhah ecosystem, enabling more flexible domain geometries and higher-fidelity simulations of vacuum, GRHD, and GRMHD systems. Together, the \nrpy framework and its \superb extension provide a unified, scalable, and extensible platform for next-generation NR modeling, suitable for deployment ranging from volunteer desktops involved in the \BHaH volunteer computing project to leadership-class supercomputers.

The remainder of this paper is structured as follows: Section~\ref{sec:formalism} reviews the BSSN formulation, gauge conditions, and tensor rescaling as implemented within \nrpy/\bhah. Section~\ref{sec:parallelization} describes the \superb/\charmpp parallelization strategy applied to \nrpy-generated code. Section~\ref{sec:results} presents validation, the head-on collision simulation, and performance results using \superbnrpy. Section~\ref{sec:concl} summarizes and discusses future work for the \nrpy framework and its \superb extension.

\section{Basic Equations}
\label{sec:formalism}

We briefly review the formalism used in \nrpy for vacuum spacetime evolution, focusing on the evolution equations and tensor rescaling. For more details, see Ref.~\cite{Ruchlin2018}. Throughout this paper we adopt geometrized units ($G=c=1$), and Latin indices ($i,j,k,\dots$) denote spatial components with repeated indices implying summation.

\subsection{Evolution equations}

Our formulation employs the covariant BSSN equations with the Lagrangian choice $\pd_t \bar \gamma = 0$ \cite{Brown2009}, adapted for curvilinear coordinates \cite{Baumgarte2013, Ruchlin2018}. The spacetime metric undergoes the standard ADM decomposition:
\be
ds^2 = -\alpha^2 dt^2 + \gamma_{ij} (dx^i + \beta^i dt)(dx^j + \beta^j dt) \ ,
\ee
where $\alpha$ is the lapse function, $\beta^i$ the shift vector, and $\gamma_{ij}$ the spatial 3-metric.

The conformal metric $\bar \gamma_{ij}$ is defined via the decomposition
\be
\gamma_{ij} = e^{4\phi} \bar \gamma_{ij} \ ,
\ee
where $\phi$ is the conformal factor. Following common practice \cite{Tichy2007, Marronetti2008}, we evolve $W = e^{-2\phi}$ for improved numerical behavior near punctures.

To effectively handle curvilinear coordinates, we introduce a time-independent reference metric $\hat \gamma_{ij}$ ($\pd_t \hat{\gamma}_{ij} = 0$), typically chosen as the flat metric in the adopted coordinate system \cite{Ruchlin2018}. The conformal metric is then decomposed relative to this reference metric as
\be
\bar \gamma_{ij} = \hat \gamma_{ij} + \varepsilon_{ij} \ ,
\ee
where $\varepsilon_{ij}$ represents the deviation from the reference metric. This decomposition is essential both for defining difference-based connection functions and for enabling the tensor rescaling described in Sec.~\ref{sec:rescaling}.

Quantities computed using the conformal metric (e.g., the covariant derivative $\bar{D}_i$ and Christoffel symbols $\bar{\Gamma}^k_{ij}$) are denoted by a bar, while those associated with the reference metric (e.g., $\hat{D}_i$ and $\hat{\Gamma}^k_{ij}$) carry a hat; both are defined with respect to the same coordinate system. The difference between their connection coefficients defines the tensor
\begin{eqnarray}
\Delta^k_{ij} &=& \bar \Gamma^k_{ij} - \hat \Gamma^k_{ij} \ ,\\[1mm]
\Delta^k &=& \bar \gamma^{ij} \Delta^k_{ij} \ .
\end{eqnarray}
A new BSSN variable, the conformal connection function $\bar \Lambda^i$, is introduced and is constrained to satisfy the algebraic constraint
\[
\mathcal{C}^i = \bar \Lambda^i - \Delta^i = 0 \, .
\]

The trace-free part of the conformal extrinsic curvature is defined as
\be
\bar A_{ij} \equiv e^{4\phi}\Bigl( K_{ij} - \frac{1}{3} \gamma_{ij} K\Bigr) \ ,
\ee
where $K_{ij}$ is the extrinsic curvature and $K = \gamma^{ij} K_{ij}$ its trace. The normal derivative operator relative to the Eulerian observer is defined as
\[
\pd_\perp = \pd_t - \Lie_\beta \, ,
\]
where $\Lie_\beta$ denotes the Lie derivative along the shift vector $\beta^i$.

The BSSN formulation evolves the metric deviation $\varepsilon_{ij}$, the trace-free conformal extrinsic curvature $\bar A_{ij}$, the conformal factor variable $W$, the trace of the extrinsic curvature $K$, and the conformal connection functions $\bar \Lambda^i$. As implemented in \nrpy, their evolution equations are:
\begin{eqnarray}
  \label{eq:metric_RHS_orig_final} 
  \partial_{\perp} \varepsilon_{ij} & = & \frac{2}{3} \bar{\gamma}_{ij} \Bigl(\alpha \bar{A}_{k}^{k} - \bar{D}_k \beta^{k}\Bigr) + 2 \hat{D}_{(i} \beta_{j)} - 2 \alpha \bar{A}_{ij} \\[1mm]
  \label{eq:Abar_RHS_orig_final} 
  \partial_{\perp} \bar{A}_{ij} & = & -\frac{2}{3} \bar{A}_{ij} \bar{D}_k \beta^{k} - 2 \alpha \bar{A}_{ik} {\bar{A}^{k}}_{j} + \alpha \bar{A}_{ij} K \nonumber \\ 
  & & + e^{-4\phi} \Bigl\{-2 \alpha \bar{D}_i \bar{D}_j \phi + 4 \alpha \bar{D}_i \phi\, \bar{D}_j \phi \nonumber \\ 
  & & \qquad + 4 \bar{D}_{(i}\alpha\, \bar{D}_{j)}\phi - \bar{D}_i\bar{D}_j\alpha + \alpha \bar{R}_{ij}\Bigr\}^{\mathrm{TF}} \\[2mm] 
  \label{eq:W_RHS_orig_final} 
  \partial_{\perp} W & = & -\frac{1}{3} W \Bigl(\bar{D}_k \beta^{k} - \alpha K\Bigr) \\[2mm]
  \label{eq:K_RHS_orig_final} 
  \partial_{\perp} K & = & \frac{1}{3} \alpha K^{2} + \alpha \bar{A}_{ij} \bar{A}^{ij} - e^{-4\phi} \Bigl(\bar{D}_i \bar{D}^{i} \alpha + 2 \bar{D}^{i} \alpha\, \bar{D}_i\phi\Bigr) \\[2mm]
  \label{eq:Lambda_RHS_orig_final} 
  \partial_{\perp} \bar{\Lambda}^{i} & = & \bar{\gamma}^{jk}\, \hat{D}_j\hat{D}_k\beta^{i} + \frac{2}{3} \Delta^{i}\, \bar{D}_j \beta^{j} + \frac{1}{3} \bar{D}^{i}\bar{D}_j\beta^{j} \nonumber \\ 
  & & - 2 \bar{A}^{ij} \Bigl(\partial_j \alpha - 6 \, \partial_j \phi\Bigr) + 2 \, \bar{A}^{jk} \Delta^i_{jk} - \frac{4}{3} \alpha\, \bar{\gamma}^{ij}\, \partial_j K 
\end{eqnarray}
Here, TF indicates that the enclosed expression is taken to be trace-free with respect to $\bar{\gamma}^{ij}$, and the conformal Ricci tensor $\bar R_{ij}$ is computed as
\begin{eqnarray}
  \bar{R}_{i j} {} = {} && - \frac{1}{2} \bar{\gamma}^{k l} \hat{D}_{k} \hat{D}_{l} \bar{\gamma}_{i j} + \bar{\gamma}_{k(i} \hat{D}_{j)} \bar{\Lambda}^{k} + \Delta^{k} \Delta_{(i j) k} \nonumber \\
  && + \bar{\gamma}^{k l} \left (2 \Delta_{k(i}^{m} \Delta_{j) m l} + \Delta_{i k}^{m} \Delta_{m j l} \right ) \ .
\end{eqnarray}
The trace-free condition $\bar{A}_{ij} \bar{\gamma}^{ij} = 0$ is enforced dynamically via the $\bar A^k_k$ term in Eq.~(\ref{eq:metric_RHS_orig_final}).

We employ the standard 1+log slicing \cite{Bona1995} and Gamma-driver shift \cite{Alcubierre2003} conditions, which evolve the lapse $\alpha$, shift $\beta^i$, and an auxiliary variable $B^i$:
\begin{eqnarray}
\label{eq:gauge_eqns}
\partial_t \alpha &=& -2 \alpha K + \beta^j \partial_j \alpha \ ,\\[1mm]
\partial_t \beta^{i} &=& B^{i} + \beta^j \partial_j \beta^i \ ,\\[1mm]
\pd_t B^{i} &=& \frac{3}{4} \partial_t \bar{\Lambda}^{i} - \eta B^{i} + \beta^j \partial_j B^i \ , \label{eq:damping_eta_final}
\end{eqnarray}
where $\eta$ is a damping parameter \cite{Schnetter2010}. Note that $\pd_t \bar \Lambda^i$ on the right-hand side of Eq.~(\ref{eq:damping_eta_final}) is substituted using Eq.~(\ref{eq:Lambda_RHS_orig_final}). These choices yield a total of 24 evolved variables. 

While the BSSN equations above describe the time evolution of the geometric variables, the constraint equations complete the description of Einstein's equations: the Hamiltonian constraint $\mathcal{H}$ and the momentum constraints $\mathcal{M}^i$, which must be satisfied on each spatial hypersurface. While mathematically preserved by the evolution equations, numerically they provide a crucial check on the accuracy of the simulation. Throughout the simulation we monitor these constraints \cite{Ruchlin2018, Baumgarte2013}:
\begin{eqnarray}
  \label{eq:Hamiltonian}
  \mathcal{H} &\equiv& \frac{2}{3} K^{2} - \bar{A}_{i j} \bar{A}^{i j} + e^{-4 \phi} \left (\bar{R} - 8 \bar{D}^{i} \phi \bar{D}_{i} \phi - 8 \bar{D}^{2} \phi \right ) = 0 \ , \label{eq:Hconstraint}\\
  \mathcal{M}^{i} &\equiv& e^{-4 \phi} \left (\hat{D}_{j} \bar{A}^{i j} + 2 \bar{A}^{k (i} \Delta^{j)}_{j k} + 6 \bar{A}^{i j} \partial_{j} \phi - \frac{2}{3} \bar{\gamma}^{i j} \partial_{j} K \right ) = 0 \ . \label{eq:Mconstraint}
\end{eqnarray}
Here, $\bar{R} = \bar{\gamma}^{i j} \bar{R}_{i j}$.

\subsection{Tensor rescaling}
\label{sec:rescaling}

Coordinate singularities intrinsic to curvilinear systems (e.g., at $r=0$ or $\theta \in \{0, \pi\}$ in spherical coordinates) can cause tensor components expressed in the coordinate basis $\hat{e}_{(i)}$ to diverge, even for physically regular fields. Numerical methods like finite differencing and interpolation require sufficiently smooth input functions; applying them directly to divergent components induces large errors and potential instabilities. Tensor rescaling addresses this challenge by expressing tensor components in a carefully chosen non-coordinate basis $\tilde{e}_{(i)}$ such that the components remain smooth and regular at these singularities \cite{Baumgarte2013, Ruchlin2018}.

Consider, for example, flat 3D space described by Cartesian coordinates $x^k=(x,y,z)$ and standard spherical coordinates $\hat x^i=(r,\theta,\phi)$ related by
\begin{eqnarray}
\label{eq:carttosph_revised_final}
x &=& r\sin\theta\cos\phi \ ,\\[1mm]
y &=& r\sin\theta\sin\phi \ ,\\[1mm]
z &=& r\cos\theta \ .
\end{eqnarray}
A tensor with Cartesian components $A^{kl}$ transforms to spherical coordinate basis components $\hat A^{ij}$ according to the standard transformation rules:
\[
\hat A^{ij} = A^{kl} \frac{\partial \hat{x}^i}{\partial x^k} \frac{\partial \hat{x}^j}{\partial x^l} \,.
\]
The flat Euclidean metric becomes the reference metric in spherical coordinates, $\hat \gamma_{ij} = \mathrm{diag}(1,\,r^2,\,r^2\sin^2\theta)$. While $\hat{\gamma}_{ij}$ itself is regular, other tensors can acquire divergent components in the spherical \emph{coordinate basis} $\hat{e}_{(i)}$ at $r=0$ or where $\sin\theta=0$. For instance, a constant Cartesian vector $V^k = (1,0,0)$ transforms to spherical coordinate components
\be
\hat V^i = (\sin\theta \cos\phi, \frac{\cos\theta \cos\phi}{r}, -\frac{\sin\phi}{r\sin\theta}) \ .
\ee
These components diverge, rendering standard numerical finite differencing, quadrature, or interpolations problematic near the singularities.

The solution involves defining a rescaled, non-coordinate basis $\tilde{e}_{(i)}$ derived from the reference metric. For a diagonal reference metric $\hat \gamma_{ij} = \mathrm{diag}(\hat \gamma_{11},\,\hat \gamma_{22},\,\hat \gamma_{33})$, we define scale factors
\[
S_{(i)} = \sqrt{\hat \gamma_{ii}}\quad (\mathrm{no\ summation\ implied}) \,,
\]
and corresponding rescaled basis vectors
\[
\tilde e_{(i)} = \frac{\hat e_{(i)}}{S_{(i)}} \,.
\]
In this orthonormal rescaled basis, the reference metric becomes the identity matrix, $\tilde \gamma_{ij} = \delta_{ij}$. Tensor components are transformed to represent them in this regular basis: contravariant components become $\tilde V^i = \hat V^i\, S_{(i)}$, and covariant components become $\tilde V_i = \hat V_i/S_{(i)}$ (no sum). For the example vector $\hat V^i$ above, the rescaled components in the tilde basis are
\[
\tilde V^i = (\sin\theta\cos\phi,\, \cos\theta\cos\phi,\, -\sin\phi) \,,
\]
which represents the same vector but using components that are regular everywhere. \nrpy automatically applies this rescaling procedure for supported curvilinear coordinate systems by using the appropriate scale factors.

\section{\charmpp parallelization of \nrpy-generated code}
\label{sec:parallelization}

The \nrpy framework generates code that operates on structured, logically rectangular grids, even when representing curvilinear coordinate systems. Spatial coordinates $x^i$ ($i=0,1,2$) correspond to $(x, y, z)$, $(r, \theta, \phi)$, or $(\rho, \phi, z)$ for Cartesian-like, spherical-like, or cylindrical-like systems, respectively. The grid spans a logical domain from $x^i_\mathrm{min}$ to $x^i_\mathrm{max}$ with $N_{x^i}$ points in each direction. Grid points are cell-centered and uniformly distributed between $x^i_\mathrm{min} + dx^i/2$ and $x^i_\mathrm{max} - dx^i/2$, where $dx^i = (x^i_\mathrm{max} - x^i_\mathrm{min})/N_{x^i}$. This cell-centering strategy inherently avoids placing grid points directly on coordinate singularities, as detailed in Ref.~\cite{Ruchlin2018}.

\begin{figure}[ht]
\centering
\begin{tikzpicture}
\def\rmin{0}           
\def\rmax{2.5}             
\def\phimin{0}           
\def\phimax{360}         
\def\Nr{6}               
\def\Nphi{12}            
\def\NG{2}
\pgfmathsetmacro{\dr}{(\rmax - \rmin)/\Nr}         
\pgfmathsetmacro{\dphi}{(\phimax - \phimin)/\Nphi} 
\foreach \i in {0,...,\Nr} {
    \draw[gray, thin] (0,0) circle ({\rmin + \i * \dr});
}
\foreach \j in {0,...,\Nphi} {
    \draw[gray, thin] (0,0) -- ({\phimin + \j * \dphi}:\rmax);
}
\foreach \i in {0,...,\numexpr\Nr-1} {
    \foreach \j in {0,...,\numexpr\Nphi-1} {
        \pgfmathsetmacro{\r}{\rmin + (\i + 0.5) * \dr}  
        \pgfmathsetmacro{\phi}{\phimin + (\j + 0.5) * \dphi}
            \fill[black] (\phi:\r) circle (1.2pt);  
    }
}
\foreach \i in {0,...,\numexpr\NG-1} {
    \foreach \j in {0,...,\numexpr\Nphi-1} {
        \pgfmathsetmacro{\r}{\rmin + (\Nr + \i + 0.5) * \dr}  
        \pgfmathsetmacro{\phi}{\phimin + (\j + 0.5) * \dphi}
            \draw[blue] (\phi:\r) circle (1.5pt);
    }
}
\foreach \i in {0,...,\numexpr\NG-1} {
    \foreach \j in {0,...,\numexpr\Nphi-1} {
        \pgfmathsetmacro{\r}{\rmin + ( \i + 0.5) * \dr}
        \pgfmathsetmacro{\phi}{\phimin + (\j + 0.5) * \dphi}
            \draw[red] (\phi:\r) circle (2.5pt);
    }
}
\foreach \i in {0,...,\numexpr\Nr+1} { 
    \foreach \j in {0,...,\numexpr\NG-1} {
        \pgfmathsetmacro{\r}{\rmin + ( \i + 0.5) * \dr}
        \pgfmathsetmacro{\phi}{180 + (\j + 0.5) * \dphi}
            \draw[red] (\phi:\r) circle (2.5pt);
        \pgfmathsetmacro{\phi}{180 + (\j - 0.5 -1) * \dphi}
            \draw[red] (\phi:\r) circle (2.5pt);
    }
}
\node[anchor=west] at (-3.75, 0.25) {$\phi=\pi$};
\node[anchor=west] at (-3.75, -0.25) {$\phi=-\pi$};
\end{tikzpicture}
\caption{\label{fig:grid}Physical grid layout in the equatorial plane for a spherical coordinate system used by \nrpy/\bhah, illustrating different ghost point types. Black dots represent interior grid points. Blue circles are outer boundary ghost points, located beyond the maximal radius ($r > r_\mathrm{max}$). Red circles are inner boundary ghost points, arising from coordinate system properties: the origin ($r=0$) and periodicity ($\phi=\pm\pi$). These inner boundary points map logically to the points they encircle, requiring specific parity conditions upon copying. Two layers of ghost points ($N_G=2$) are shown.}
\end{figure}
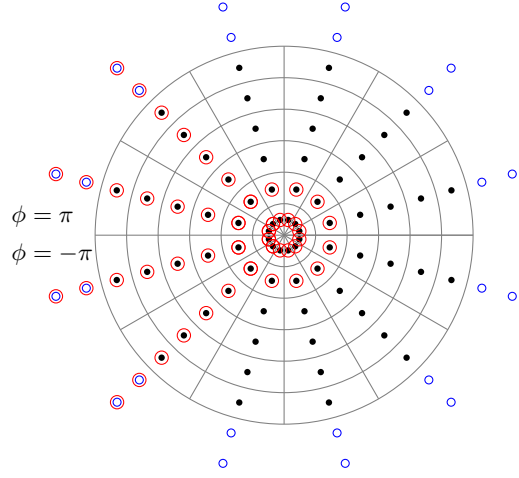

Spatial derivatives are computed using finite differencing of selectable order $N_\mathrm{FD}$ (typically 2, 4, 6, 8, 10). Most derivatives employ centered stencils, requiring $N_G = N_{\mathrm{FD}}/2$ layers of ghost points surrounding the computational domain to compute derivatives near the boundaries. By contrast, shift--advection derivatives (terms involving $\beta^i\partial_i$) are computed using upwind finite differences, which require $N_G = N_{\mathrm{FD}}/2 + 1$ ghost layers on the appropriate side. The nature of these ghost points depends on the coordinate system and the location on the numerical grid. Figure~\ref{fig:grid} visualizes a spherical grid, illustrating two distinct types:
\begin{itemize}
    \item \textbf{Outer boundary points} (blue circles in Fig.~\ref{fig:grid}) lie beyond the physical boundary of the domain (e.g., near $r=r_\mathrm{max}$). These are filled using physical boundary conditions, such as Sommerfeld radiation conditions \cite{Assumpcao2022} or extrapolation.
    \item \textbf{Inner boundary points} (red circles in Fig.~\ref{fig:grid}) arise from coordinate singularities (e.g., at $r=0$) or periodicity (e.g., at $\phi=\pm\pi$). These points logically map to interior grid points or outer boundary points elsewhere on the grid (indicated by the points they encircle in Fig.~\ref{fig:grid}). They are filled using values from the corresponding points, applying appropriate parity transformations based on the tensor type and coordinate symmetries, a process handled automatically by \nrpy \cite{Ruchlin2018}.
\end{itemize}
In the original single-node \nrpy implementation, filling both types of ghost points involves only local data access or calculation.

\begin{figure}
\centering
\begin{tikzpicture}
    \def\xmin{0}           
    \def\xmax{6}           
    \def\ymin{0}           
    \def\ymax{3}           
    \def\Nx{12}             
    \def\Ny{6}             
    \def\NG{2}              
    \pgfmathsetmacro{\dx}{(\xmax - \xmin)/\Nx}         
    \pgfmathsetmacro{\dy}{(\ymax - \ymin)/\Ny}         
    \node[rotate=90,anchor=west] at (\xmin - 2.5*\dx, \ymin- 1*\dy) {$r=0$};
    \node[rotate=90,anchor=west] at (\xmin - 2.5*\dx, \ymax- 1*\dy) {$r=r_\mathrm{max}$};
    \node[anchor=west] at (\xmin - 1*\dx, \ymin- 2.5*\dy) {$\phi=-\pi$};
    \node[anchor=west] at (\xmax- 1*\dx , \ymin- 2.5*\dy) {$\phi=\pi$};
    \foreach \i in {0,...,\Nx} {
        \draw[gray, thin] (\xmin + \i * \dx, \ymin) -- (\xmin + \i * \dx, \ymax);
    }
    \foreach \i in {3,6,...,\numexpr\Nx-3} {
        \draw[black, line width=1.2pt, dashed] (\xmin + \i * \dx, \ymin - 2 * \dy) -- (\xmin + \i * \dx, \ymax + 2 * \dy);
    }
    \foreach \j in {0,...,\Ny} {
        \draw[gray, thin] (\xmin, \ymin + \j * \dy) -- (\xmax, \ymin + \j * \dy);
    }
    \foreach \j in {3,...,\numexpr\Ny-3} {
        \draw[black, line width=1.2pt, dashed] (\xmin - 2 * \dx, \ymin + \j * \dy) -- (\xmax + 2 * \dx, \ymin + \j * \dy);
    }
    \foreach \i in {0,...,\numexpr\Nx-1} {
        \foreach \j in {0,...,\numexpr\Ny-1} {
            \pgfmathsetmacro{\x}{\xmin + (\i + 0.5) * \dx}  
            \pgfmathsetmacro{\y}{\ymin + (\j + 0.5) * \dy}  
            \fill[black] (\x, \y) circle (1.2pt);  
        }
    }

    \draw[green] (\xmin + 1.5*\dx , \ymax + 1.5*\dy) circle (2.5pt);
    \draw[green] (\xmin + 1.5*\dx , \ymax + 0.5*\dy) circle (2.5pt);
    \draw[green] (\xmin + 1.5*\dx , \ymax - 0.5*\dy) circle (2.5pt);
    \draw[green] (\xmin + 1.5*\dx , \ymax - 1.5*\dy) circle (2.5pt);
    \draw[green] (\xmin + 1.5*\dx , \ymax - 2.5*\dy) circle (2.5pt);
    \draw[green] (\xmin + 2.5*\dx , \ymax + 1.5*\dy) circle (2.5pt);
    \draw[green] (\xmin + 2.5*\dx , \ymax + 0.5*\dy) circle (2.5pt);
    \draw[green] (\xmin + 2.5*\dx , \ymax - 0.5*\dy) circle (2.5pt);
    \draw[green] (\xmin + 2.5*\dx , \ymax - 1.5*\dy) circle (2.5pt);
    \draw[green] (\xmin + 2.5*\dx , \ymax - 2.5*\dy) circle (2.5pt);
    \draw[green] (\xmin + 6.5*\dx , \ymax + 1.5*\dy) circle (2.5pt);
    \draw[green] (\xmin + 6.5*\dx , \ymax + 0.5*\dy) circle (2.5pt);
    \draw[green] (\xmin + 6.5*\dx , \ymax - 0.5*\dy) circle (2.5pt);
    \draw[green] (\xmin + 6.5*\dx , \ymax - 1.5*\dy) circle (2.5pt);
    \draw[green] (\xmin + 6.5*\dx , \ymax - 2.5*\dy) circle (2.5pt);
    \draw[green] (\xmin + 7.5*\dx , \ymax + 1.5*\dy) circle (2.5pt);
    \draw[green] (\xmin + 7.5*\dx , \ymax + 0.5*\dy) circle (2.5pt);
    \draw[green] (\xmin + 7.5*\dx , \ymax - 0.5*\dy) circle (2.5pt);
    \draw[green] (\xmin + 7.5*\dx , \ymax - 1.5*\dy) circle (2.5pt);
    \draw[green] (\xmin + 7.5*\dx , \ymax - 2.5*\dy) circle (2.5pt);
    \draw[green] (\xmin + 1.5*\dx , \ymax - 3.5*\dy) circle (2.5pt);
    \draw[green] (\xmin + 2.5*\dx , \ymax - 3.5*\dy) circle (2.5pt);
    \draw[green] (\xmin + 3.5*\dx , \ymax - 3.5*\dy) circle (2.5pt);
    \draw[green] (\xmin + 4.5*\dx , \ymax - 3.5*\dy) circle (2.5pt);
    \draw[green] (\xmin + 5.5*\dx , \ymax - 3.5*\dy) circle (2.5pt);
    \draw[green] (\xmin + 6.5*\dx , \ymax - 3.5*\dy) circle (2.5pt);
    \draw[green] (\xmin + 7.5*\dx , \ymax - 3.5*\dy) circle (2.5pt);
    \draw[green] (\xmin + 1.5*\dx , \ymax - 4.5*\dy) circle (2.5pt);
    \draw[green] (\xmin + 2.5*\dx , \ymax - 4.5*\dy) circle (2.5pt);
    \draw[green] (\xmin + 3.5*\dx , \ymax - 4.5*\dy) circle (2.5pt);
    \draw[green] (\xmin + 4.5*\dx , \ymax - 4.5*\dy) circle (2.5pt);
    \draw[green] (\xmin + 5.5*\dx , \ymax - 4.5*\dy) circle (2.5pt);
    \draw[green] (\xmin + 6.5*\dx , \ymax - 4.5*\dy) circle (2.5pt);
    \draw[green] (\xmin + 7.5*\dx , \ymax - 4.5*\dy) circle (2.5pt);

    \foreach \i in {0,...,\numexpr\NG-1} {
        \foreach \j in {-2,...,\numexpr\Ny+1} {
            \pgfmathsetmacro{\x}{\xmin + (\Nx + \i + 0.5) * \dx}  
            \pgfmathsetmacro{\y}{\ymin + (\j + 0.5) * \dy}  
            \draw[red] (\x, \y) circle (2.5pt);
            \pgfmathsetmacro{\x}{\xmin + (-\i - 0.5) * \dx}  
            \pgfmathsetmacro{\y}{\ymin + (\j + 0.5) * \dy}  
            \draw[red] (\x, \y) circle (2.5pt);
        }
    }
    \foreach \i in {0,...,\numexpr\Nx-1} {
        \foreach \j in {0,...,\numexpr\NG-1} {
            \pgfmathsetmacro{\x}{\xmin + (\i + 0.5) * \dx}
            \pgfmathsetmacro{\y}{\ymin + (\Ny + \j + 0.5) * \dy}
            \draw[blue] (\x, \y) circle (1.5pt);
            \pgfmathsetmacro{\x}{\xmin + (\i + 0.5) * \dx}
            \pgfmathsetmacro{\y}{\ymin + (- \j - 0.5) * \dy}
            \draw[red] (\x, \y) circle (2.5pt);
        }
    }
    \fill[fill opacity=0.1, gray] (\xmin+3*\dx, \ymax) rectangle (\xmin+6*\dx, \ymax-3*\dy);
\end{tikzpicture}
\caption{\label{fig:grid_divide}Schematic of the parallel decomposition strategy. The spherical grid from Fig.~\ref{fig:grid} is shown mapped to its logically rectangular domain $(r, \phi)$ and partitioned into chares (indicated by dashed lines). Black dots are interior points; blue circles are outer boundary ghost points ($r>r_\mathrm{max}$); red circles are inner boundary ghost points ($r=0$ mapping and $\phi=\pm\pi$ periodicity). The shaded region highlights one example chare. This partitioning introduces \textbf{neighbor ghost points} (green circles), which map to grid points lying on neighboring chares. Communication via message passing is required to fill some inner boundary points and all neighbor points correctly.}
\end{figure}
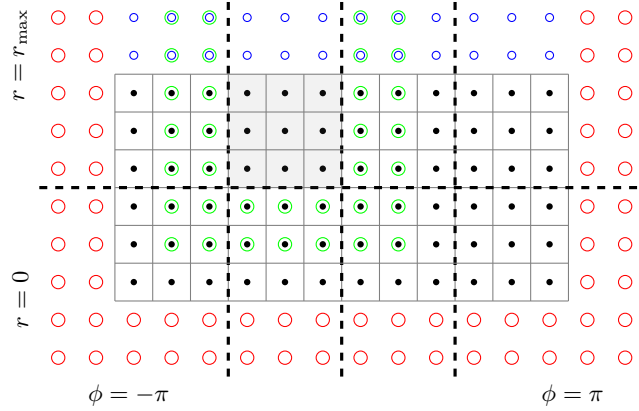

To enable distributed-memory parallelism using \charmpp, the \superb extension partitions the global logically rectangular computational grid into smaller rectangular blocks, known as `chares'. Figure~\ref{fig:grid_divide} illustrates this partitioning concept for the spherical grid depicted logically via its $(r, \phi)$ coordinates. The grid is divided along each logical coordinate direction $x^i$ into $N_{\mathrm{chare}^i}$ segments, resulting in a total of $N_{\mathrm{chare}^0} \times N_{\mathrm{chare}^1} \times N_{\mathrm{chare}^2}$ chares (Fig.~\ref{fig:grid_divide} shows an example with $N_{\mathrm{chare}^r}=2, N_{\mathrm{chare}^\phi}=4$). Each chare is responsible for managing a subgrid containing $\prod_{i=0}^{2} (N_{x^i} / N_{\mathrm{chare}^i})$ interior points, ensuring this size remains larger than the required ghost zone width $N_G$.

This domain decomposition necessitates communication between chares. Since finite differencing stencils require data from neighboring grid points, these points may now reside on different chares (and thus different processing units or nodes). Filling the ghost points for each chare now involves managing three distinct scenarios, building upon the classifications illustrated in Figure~\ref{fig:grid_divide}:
\begin{enumerate}
    \item \textbf{Outer boundary points:} Chares located at the global physical boundary (e.g., the top edge at $r=r_\mathrm{max}$ in Fig.~\ref{fig:grid_divide}) fill their respective outer ghost points (blue circles) locally using the prescribed physical boundary conditions (e.g., Sommerfeld), identical to the single-node case. No inter-chare communication is needed for these points.
    \item \textbf{Inner boundary points:} These points (red circles at $r=0$ and $\phi=\pm\pi$ edges in Fig.~\ref{fig:grid_divide}) still logically map to interior or outer boundary points elsewhere on the global grid. If the source point for the mapping resides within the \emph{same} chare, the data is copied locally with appropriate parity transformations, as before. However, if the source point lies on a \emph{different} chare (which might be non-adjacent in the logical grid partitioning, especially for mappings near $r=0$), \superbnrpy generates code implementing direct point-to-point communication using \charmpp messages to retrieve the required data. The underlying mapping logic from \nrpy is used to identify the coordinates of the target source point and determine its owning chare. This is optimized by having each chare, at initial setup, compute a list of points required by other chares and sending only the required data directly to each chare.
    
    \item \textbf{Neighbor ghost points:} These points (green circles for the shaded chare in Fig.~\ref{fig:grid_divide}) are interior, inner boundary or outer boundary points of one chare that are needed to fill the ghost zones of an adjacent chare due to the finite difference stencil overlapping the boundary between chares. These points constitute the ``halo'' region and must be populated via inter-chare communication. This is optimized by sending face data, including ghost points, sequentially in the east-west, then north-south, then top-bottom directions. Correct data for the ghost points in chares sharing an edge or a corner are thus automatically updated without direct communication with these adjacent chares.
\end{enumerate}

To ensure correct data is available for derivatives, the boundary conditions and communication must be performed in a specific order within each Runge-Kutta time step substage:
\begin{enumerate}
    \item Apply outer boundary conditions locally on relevant chares.
    \item Apply inner boundary conditions, involving local copies (intra-chare mappings) and point-to-point inter-chare communication (remote mappings).
    \item Exchange neighbor ghost point data (halo exchange) between adjacent chares. This step typically uses the standard technique of communicating face data sequentially (e.g., east-west, then north-south, then top-bottom) to correctly update face, edge, and corner neighbor data.
\end{enumerate}
This sequence guarantees that all required ghost point data is valid before finite differences are computed for the interior points of each chare.

This communication pattern is executed after each substage of the chosen time integrator (e.g., SSPRK3~\cite{GottliebShuTadmorSSPRK}). We use the \ckio library within \charmpp for efficient parallel I/O of diagnostic data and integrate \charmpp's native checkpointing capabilities for enhanced fault tolerance.

\section{Results}
\label{sec:results}

The \nrpy framework \cite{Ruchlin2018}, upon which \superbnrpy is built, has undergone rigorous validation. Specifically, \nrpy-generated codes, including \bhah, have achieved round-off level agreement with established NR codes in spherical coordinates \cite{Baumgarte2013}, produced results consistent with other \et modules for BBH evolutions in Cartesian coordinates \cite{Etienne2024}, and exhibited expected convergence rates for constraint violations across various coordinate systems. Further, the \nrpy-based \bhah code accurately extracts ringdown waveforms via the Newman--Penrose scalar $\psi_4 = \ddot h_{+} - i\,\ddot h_{\times}$, matching analytical quasinormal-mode frequencies and exponential amplitude damping rates \cite{Ruchlin2018} for a dominant mode. Since \superbnrpy reuses the core physics and infrastructure modules from \nrpy also employed by \bhah, it directly inherits this robustly validated foundation.

\begin{figure}[!ht]
\centering
\includegraphics{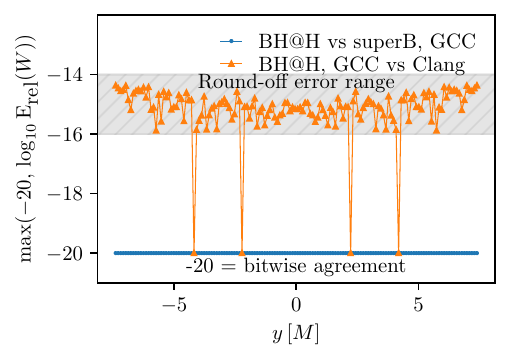}
\caption{\label{fig:agreement}Demonstration of bitwise agreement for the evolved variable $W$ along the $y$-axis at the end of an axisymmetric head-on BH collision simulation in cylindrical-like coordinates on a 16-core desktop comparing \bhah (\openmp) and \superb (\charmpp, partitioned with $N_{\mathrm{chare}^i}= \{18,2,1\}$). When using the same compiler, the results are identical (solid circles, difference is zero). For comparison, using \bhah compiled with different compilers results in expected round-off-level differences (solid triangles).}
\end{figure}

The primary contribution of this work is the \charmpp-based distributed-memory parallelization enabled by \superb. We first verified its correctness by comparing the computed values of evolved variables, such as the conformal factor variable $W \equiv e^{-2\phi}$, against those from the standard \openmp-based \bhah code generated by \nrpy for a small grid simulation of a head-on BH collision in axisymmetry using cylindrical-like coordinates on a 16-core desktop. When compiled with the same compiler and run with identical parameters on a single node, the \charmpp code generated by \superbnrpy produces bit-identical results to the \openmp code, as shown in Fig.~\ref{fig:agreement}. As part of this work, we also extended the decomposition of the Weyl scalar $\psi_4$ using spin-weighted spherical harmonics ($_{-2}Y_{\ell m}$) to properly handle cylindrical-like coordinates, enabling waveform extraction in these systems within the \superbnrpy framework. 

To demonstrate the capabilities and performance of \superbnrpy, particularly for simulations benefiting from curvilinear coordinates and parallelism, we present two main sets of results. First, we analyze a head-on BH collision, validating waveform extraction against QNM predictions and examining constraint violations, including the impact of the SSL technique \cite{Etienne2024} (which is implemented within the shared \nrpy/\bhah infrastructure and thus available in \superbnrpy). Second, we present performance benchmarks, including single-node comparisons to the \openmp baseline and multi-node strong scaling tests on an HPC cluster.

\subsection{Black hole head-on collision}
\label{sec:results_headon}

We simulate the head‐on collision of two equal‐mass, non‐spinning puncture BHs with bare masses 
$m_1 = m_2 = 0.5\,M$, initially at rest on the $z$‐axis with a coordinate separation of $10\,M$.\footnote{This example can be easily reproduced. First, install \nrpy by running \texttt{pip install nrpy}. (Further details can be found in the project's `README.md' file at \url{https://github.com/nrpy/nrpy}.) Next, install \charmpp by following the official manual. Finally, run \texttt{python3 -m nrpy.examples.superB\_blackhole\_spectroscopy --paper} to generate the example code and follow the instructions displayed on the terminal for compiling and running.} This axisymmetric scenario benefits from the use of cylindrical-like coordinates generated via \nrpy/\bhah, which naturally concentrate resolution near the collision axis ($\rho=0$) and the origin. These coordinates are defined by applying a sinh rescaling to the standard cylindrical coordinates $(\rho, z)$:
\be
x'(x) = x_\mathrm{max} \frac{\sinh(x/ w)}{\sinh(1/w)} \ ,
\ee
where $x$ represents the original coordinate ($\rho$ or $z$), $x_\mathrm{max}$ is the outer boundary of the domain extent in that coordinate, and $w=0.2$ is a parameter controlling the concentration of points near the origin ($x=0$). The computational grid domain extends from $\rho_\mathrm{min}=0$ to $\rho_\mathrm{max} = 300\,M$ and from $z_\mathrm{min} = -300\,M$ to $z_\mathrm{max} = 300\,M$. The simulation runs until a final time of $t=450\,M$.

Initial data are generated using \twopunctures \cite{Ansorg2004}. The initial spatial slice is specified to be conformally flat and maximal ($K=0$). The evolution employs the standard moving puncture gauge conditions (1+log slicing for $\alpha$ and the Gamma-driver shift condition for $\beta^i$). This mismatch between initial data ($K=0$) and gauge condition ($1+\log$) generates a known initial pulse of spurious (``junk'') radiation \cite{Alcubierre2003_2, Alcubierre2005, Etienne2014}. We use eighth-order spatial finite differencing ($N_\mathrm{FD} = 8$), eighth-order Sommerfeld radiation boundary conditions, SSPRK3 time integration with a Courant-Friedrichs-Lewy (CFL) factor of 0.5, and Gamma-driver damping parameter $\eta = 2.0$. Kreiss-Oliger dissipation is applied with strength $k = 0.99$ for gauge variables ($\alpha, \beta^i, B^i$) and $k = 0.3$ for metric variables, using ninth-order finite-differencing for the dissipation operators. We compare runs performed with and without the SSL technique enabled.

The baseline grid resolution is $N_{(\rho, \phi, z)} = \{800, 2, 2400\}$. Due to axisymmetry, only two points are needed in the $\phi$ direction. We perform runs at low (1x), medium (1.25x), and high (1.5x) resolutions, achieved by scaling $N_\rho$ and $N_z$ proportionally.

\begin{figure}[!ht]
\centering
\includegraphics{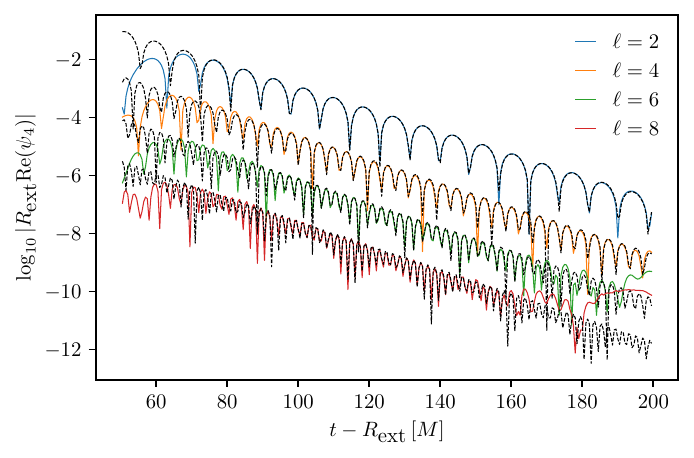}
\caption{\label{fig:psi4_fit}Head-on collision of two equal-mass, non-spinning BHs from an initial separation of $10\,M$. Real part of the extracted Weyl scalar $\psi_4$, decomposed into spin-weighted spherical harmonics $_{-2}Y_{\ell m}$, for the $m=0$ modes ($\ell = 2, 4, 6, 8$). Waveforms are extracted at radius $R_\mathrm{ext}=80\,M$ from the high-resolution simulation with SSL enabled (solid colored lines). These are compared against analytical fits (dashed black lines) using the fundamental quasi-normal mode (QNM) frequencies and damping rates for the final Schwarzschild BH \cite{Stein:2019mop}.  Deviations near the waveform peak are primarily caused by overtones \cite{Anninos1993}, which are present in the simulation but omitted from our fundamental-mode-only fit. The late-time deviations occur as the physical signal decays below the simulation's numerical noise floor.}
\end{figure}

Gravitational waveforms are extracted by decomposing the Weyl scalar $\psi_4$ into spin-weight -2 spherical harmonics up to $\ell=8$ at various radii. To validate the physical accuracy of the evolution and extraction, we compare the ringdown portion of the dominant modes with analytical predictions from BH perturbation theory. Figure~\ref{fig:psi4_fit} shows the real part of the $m=0$ modes ($\ell=2, 4, 6, 8$) extracted at $R_\mathrm{ext}=80\,M$ for the high-resolution run with SSL enabled, plotted against fits using the known fundamental quasi-normal mode (QNM) frequencies ($\omega_{\ell 0 0}$) and damping rates ($\alpha_{\ell 0 0}$) for a Schwarzschild BH \cite{Stein:2019mop}. The amplitude $A_f$ and phase $\phi_f$ are fitted between retarded times $t - R_\textrm{ext} = 100\,M$ and $130\,M$. As seen in the figure, the numerical waveforms show excellent agreement with the analytical QNM predictions for both frequency and damping during the ringdown phase. Minor deviations near the waveform peak are expected due to the presence of decaying overtones \cite{Anninos1993}, which are not included in our simple fundamental mode fit. The late-time deviations occur as the physical signal decays below the simulation's numerical noise floor.

\begin{figure*}[!ht]
\centering
\includegraphics{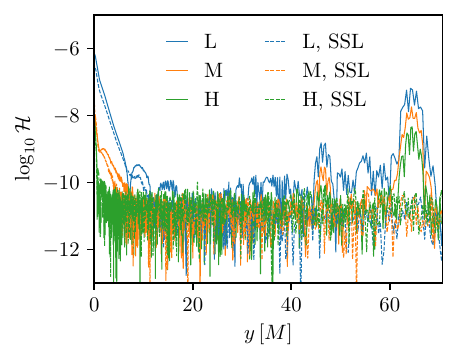}
\includegraphics{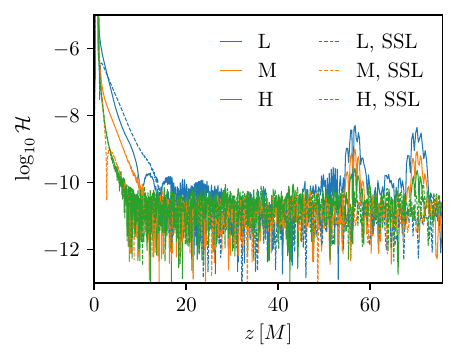}
\includegraphics{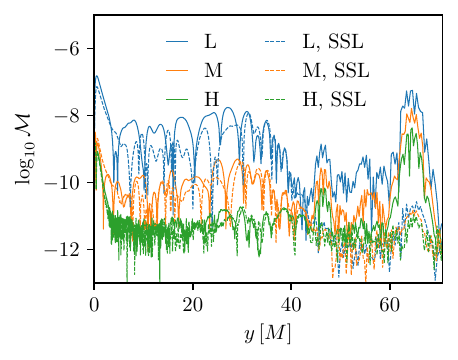}
\includegraphics{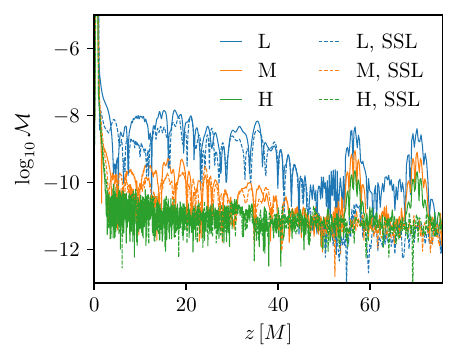}
\caption{\label{fig:constraints}Head-on collision of two BHs from a separation of $10\,M$. Logarithm of Hamiltonian, $\mathcal{H}$, and momentum, $\mathcal{M}\equiv \sqrt{\mathcal{M}_i \mathcal{M}^i}$, constraint violations (Eqs.~(\ref{eq:Hconstraint}) and~(\ref{eq:Mconstraint})) along the $y$-axis ($\rho$-axis at $z=0$) (left panels) and $z$-axis ($\rho=0$) (right panels) at $t=50\,M$. Results are shown for low (L), medium (M), and high (H) resolutions, comparing runs with SSL (dashed lines) and without SSL (solid lines).}
\end{figure*}

We also examine the evolution of constraint violations, which serve as a measure of numerical accuracy, and test the impact of the SSL technique. Figure~\ref{fig:constraints} presents the logarithm of the Hamiltonian and momentum constraint violations along the $y$-axis ($\rho$-axis at $z=0$) and $z$-axis ($\rho=0$) at $t=50\,M$ for the three resolutions, comparing runs with and without SSL. The SSL technique significantly reduces constraint violations---particularly the transient spikes associated with the initial gauge dynamics---by about an order of magnitude, consistent with the findings in Ref.~\cite{Etienne2024}. Further, the constraints exhibit the expected convergence towards zero with increasing resolution for both sets of runs.

\subsection{Performance tests}

\subsubsection{Single node}
\label{sec:single_node}

We modify the test case from that in Sec.~\ref{sec:results_headon}, by setting the BHs initially at a coordinate separation of $1\,M$ and sharing a common horizon similar to Sec. V B 3 of Ref.~\cite{Ruchlin2018}, while the outer boundary is at $7.5\,M$. This configuration of the BHs at close separation and merged are well-suited for spherical-like coordinates that present a more stringent test than cylindrical-like and Cartesian-like coordinates for parallel scaling due to more complex communication patterns. We increase the resolution in the $\phi$ direction to create a genuinely 3D problem.  The test setup uses a grid size of $N_{x^i} = \{N_r, N_\theta, N_\phi\} = \{72, 12, 24\}$. We employ fourth-order finite differencing ($N_\mathrm{FD} = 4$), requiring $N_G=N_{\mathrm{FD}}/2+1=3$ ghost zones to handle the upwinded shift advection terms.

Recall that the original \nrpy framework generates code parallelized using \openmp for intra-node shared-memory execution, which typically incurs relatively low overhead. In contrast, our \superbnrpy extension generates \charmpp code that partitions the grid into chares. This approach introduces overhead, primarily from the explicit management of tasks and the inter-chare communication required to fill ghost layers ($N_G=3$ in this test), even when running on a single node.

We first compare the performance of the \charmpp parallelization against the original \openmp version on a single node. This comparison quantifies the baseline overhead associated with the distributed-memory approach, which is relevant for users running on desktops or single HPC nodes (e.g., for volunteer computing via \BHaH). We perform these tests on a 16-core desktop machine equipped with an AMD Ryzen 9 3950X processor.

We run the 3D test case described above for $7.5\,M$ ($8428$ iterations) with diagnostics output disabled. Table~\ref{tab:single_node_perf} summarizes the wall-clock times for code generated by \superbnrpy, using either the shared-memory SMP \charmpp build or the \charmpp MPI build, both with partitioning $N_{\mathrm{chare}^i}= \{4,2,2\}$, and code generated by \nrpy targeting its standard \openmp backend.

\begin{table}[htbp]
\centering
\caption{Wall-clock time (seconds) for single desktop node test case.}
\label{tab:single_node_perf}
\begin{tabular}{lc}
Version & Wall-clock time [s] \\
\hline
\nrpy (\openmp backend, 32 threads) & 37.2 \\
\superbnrpy (\charmpp SMP, 16 cores) & 62.0 \\
\superbnrpy (\charmpp MPI, 16 cores) & 82.1 \\
\end{tabular}
\end{table}

As expected, the original \nrpy-generated \openmp version executes fastest on a single node due to its lower overhead compared to the explicit tasking and message passing inherent in the \charmpp versions. The \charmpp SMP build, which uses memory copies for communication between cores on the same node, is slower due to the overhead of managing explicit ghost zone data transfers and task scheduling. The \charmpp MPI build incurs the highest overhead on a single node due to the involvement of the network protocol stack, even when communication occurs via the loopback interface.

\begin{figure*}[!ht]
\centering
\includegraphics[width=0.8\linewidth]{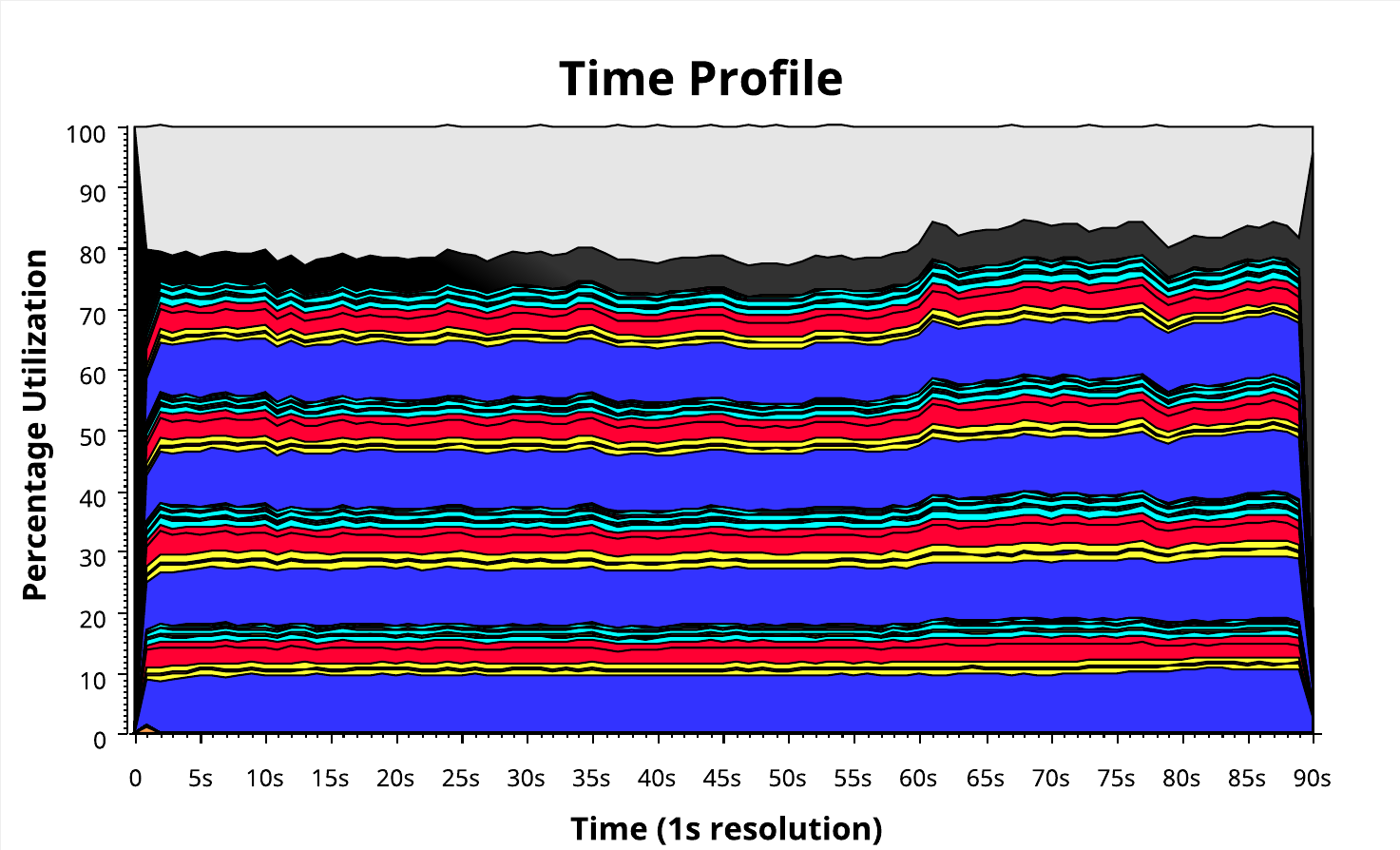}
\caption{\label{fig:projections_desktop}
(Color online.) Time profile view generated by the \projections{} analysis tool \cite{Kale2006} for the 3D BH vacuum test case running on a single 16-core desktop node using the \superbnrpy-generated code (\charmpp{} MPI build, $N_{\mathrm{chare}}=16$). The plot shows average core utilization over the total execution time. The repeated sequence of colored blocks corresponds to the substages of the RK4 time integrator: blue segments represent right-hand-side evaluation and boundary condition application; red segments correspond to the RK substage update step; yellow segments indicate communication for inner boundary ghost points; cyan segments show communication for neighbor ghost points (halo exchange); gray regions represent idle time (approximately $20\%$ in this run); and black indicates overhead associated with the \charmpp{} runtime system.}
\end{figure*}

To gain further insight into the performance characteristics of the \charmpp version, we made use of the \projections analysis tool \cite{Kale2006}. Figure~\ref{fig:projections_desktop} shows the time profile visualization for the MPI build run on the single 16-core desktop node, indicating average core utilization. The colored blocks, repeated four times, correspond to the computations for the four RK4 substages.\footnote{While the head-on collision used SSPRK3, these performance benchmarks employed RK4.} The blue regions correspond to evaluation of the right-hand-side of the RK substages and applying boundary conditions, the red regions correspond to the RK substages update, the yellow regions correspond to sending and processing inner boundary points residing on other chares' grids and the cyan regions correspond to exchanging neighbor ghosts points. The gray region corresponds to idle time, which is approximately 20\%, and the black region to the \charmpp runtime overhead. The idle time arises partly from load imbalance between chares and synchronization waits during the communication phases (halo exchange, point-to-point messages for inner boundaries). Future work involving multi-patch grids may offer better load balancing and potentially improve performance by overlapping computation and communication more effectively, leveraging \charmpp's task-based asynchronous execution model.

\subsubsection{Strong scaling on multiple nodes}

To assess the distributed-memory scalability---the primary objective of developing \superbnrpy---we performed strong scaling tests. While both strong and weak scaling are valuable performance metrics, their relevance is application-dependent. We focus on strong scaling as it is the most critical benchmark for the scientific problems \superb is designed to solve.

Weak scaling, where the problem size per core is held constant, is indispensable for applications where the required resolution scales with the domain volume, such as simulations in which e.g., GR(M)HD turbulence plays an important role. In contrast, \superb is engineered to tackle challenges in \emph{vacuum} binary black hole evolution, such as extreme mass-ratio inspirals, high spins, or relativistic scattering. In these scenarios, the primary difficulty is not an expanding domain but the need to resolve disparate physical scales in highly localized regions (e.g., near the smaller black hole's horizon).

The most computationally efficient strategy for such problems involves concentrating resources where they are most needed. This is typically achieved with multi-patch grid structures, where additional, finer grids are placed around features of interest. This approach keeps the total problem size (i.e., the total number of grid points across all patches) relatively fixed, even as the physical parameters become more extreme. Thus, the critical performance question is how efficiently the code can solve a large, fixed-size problem as more computational resources are allocated. This is precisely what a strong scaling analysis measures.

To this end, we conducted our strong scaling tests on the WindRiver HPC cluster at Idaho National Laboratory. This cluster consists of nodes equipped with dual-socket Intel Xeon Platinum 8480+ ``Sapphire Rapids'' CPUs, providing 56 cores per socket, for a total of 112 physical cores per node. To ensure sufficient computational work per core and that our test was representative of future production runs, we used a large 3D grid with $N_{x^i} = \{N_r, N_\theta, N_\phi\} = \{1008, 168, 336\}$, totaling approximately $5.7 \times 10^7$ grid points. The simulation was run for a very short duration ($0.0001\,M$, corresponding to 124 iterations) with diagnostic output disabled to focus purely on the computational scaling.

We executed the runs on 1, 8, 27, and 64 nodes, corresponding to 112, 896, 3024, and 7168 total cores, respectively. In each run, the total number of \charmpp chares was set equal to the total number of cores. The partitioning across the logical grid dimensions ($N_{\mathrm{chare}^r}, N_{\mathrm{chare}^\theta}, N_{\mathrm{chare}^\phi}$) was adjusted for each run to keep the aspect ratio of the chares constant while matching the total number of chares to the core count.

\begin{figure}[!ht]
\centering
\includegraphics{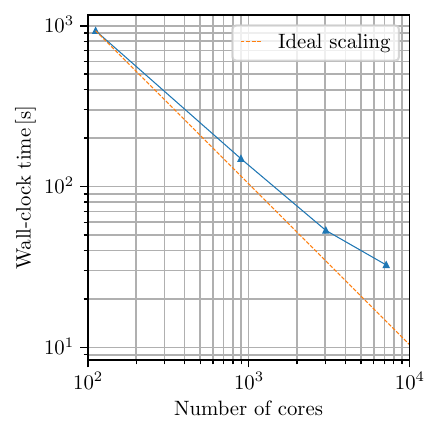}
\caption{\label{fig:strong_scaling} Strong scaling performance of \superbnrpy for the large 3D vacuum BH test case on the WindRiver HPC cluster (wall-clock time vs. number of cores). The number of chares is set equal to the number of cores. The dashed orange line indicates ideal scaling (slope -1). The speedup of the 64-node (7168 core) run relative to the 1-node (112 core) \superbnrpy run is approximately 29x. Compared to the original single-node \nrpy/\openmp code, the speedup at 64 nodes is approximately 45x.}
\end{figure}

Figure~\ref{fig:strong_scaling} presents the resulting wall-clock time versus the total number of cores used. The plot demonstrates performance scaling that is close to ideal (indicated by the dashed line with slope -1) up to approximately 1000 cores. Good scaling continues through the 64-node (7168-core) run. The speedup achieved by the 64-node \superbnrpy run relative to its 1-node counterpart (both using the \charmpp code) is approximately 29x. More significantly, comparing the 64-node \superbnrpy run (wall time $\approx$ 33 s) to the original single-node \nrpy-generated \openmp code run on one node of WindRiver (using all 112 cores, wall time $\approx$ 1462 s for this large problem), the \superbnrpy code achieves a speedup of approximately 45x. This result demonstrates the substantial performance advantage enabled by the distributed-memory parallelization for executing large-scale NR simulations on HPC clusters. Because the \superb (\charmpp with MPI build) code outperforms the original \openmp version on a single WindRiver node, we concluded that pursuing a hybrid \openmp-MPI approach (\openmp within a node and MPI across nodes) is not worthwhile.   

\section{Conclusions}
\label{sec:concl}

The rapidly evolving field of gravitational-wave astronomy demands increasingly sophisticated and computationally intensive numerical relativity simulations. While the \nrpy framework offers powerful tools for generating optimized C/C++ code for NR---particularly \bhah for curvilinear-coordinate simulations---its prior restriction to single-node \openmp parallelism limited its applicability to large-scale astrophysical systems.

In this paper, we introduced \superb, a major extension to the \nrpy ecosystem designed to overcome this limitation. \superb leverages the task-based parallelism features of the \charmpp programming system to automatically generate scalable, distributed-memory C++ code directly from existing \nrpy/\bhah physics modules. The resulting integrated framework, \superbnrpy, partitions the logically rectangular grids employed by \nrpy into computational tasks (chares) distributed across multiple nodes and cores. Crucially, \superb automatically generates the code to handle the requisite communication for ghost zone updates, including those associated with physical boundaries, coordinate singularities, and inter-task adjacencies, using local transfers, point-to-point messaging, and halo exchanges.

We validated the correctness and capabilities of the \superbnrpy-generated code through several tests. We demonstrated bit-identical numerical results compared to the original \openmp code when run on a single node with identical parameters and compiler, and performed a physically relevant head-on binary black hole collision using cylindrical-like coordinates. The extracted gravitational waveforms showed excellent agreement with analytical quasi-normal mode predictions (up to $\ell=8$) during the ringdown phase. We also demonstrated seamless compatibility with recent algorithmic enhancements implemented within the shared \nrpy framework, such as the SSL technique \cite{Etienne2024}, which effectively suppressed constraint violations.

Performance benchmarks confirmed the excellent scalability of the \superbnrpy approach. While the original \openmp code remains efficient for smaller problems on a single 16-core node, the \charmpp-based code generated by \superbnrpy exhibits strong scaling across distributed-memory HPC systems. On the high-core-count WindRiver cluster, we measured a speedup of approximately 45x for a large 3D vacuum evolution running on 64 nodes (7168 cores) compared to the optimized single-node \openmp code baseline. This demonstrates the effectiveness of \superbnrpy in enabling large-scale simulations.

The \superb extension provides a robust and scalable foundation for the entire \nrpy ecosystem. Its immediate impact includes enabling \nrpy-generated codes for problems requiring significantly larger computational resources, such as exploring challenging binary parameter spaces (e.g., high spins or large mass ratios) or performing rapid follow-up simulation campaigns for significant gravitational-wave detections. The utility of this infrastructure was highlighted by the rapid integration (approx. two weeks) of \charmpp parallelism into the \nrpy-generated \groovy GRHD code~\cite{Jacques:2024pxh}. This minimal effort was the direct result of \superb being built specifically to interface with the \bhah framework. \groovy was an existing code built within \bhah, and \superb was engineered to provide a \charmpp-based distributed-memory layer for such codes. Therefore, the core evolution equations and infrastructure were already compatible. The short integration period primarily consisted of implementing communication routines for non-evolved quantities, as is typically required for GRHD codes. We therefore anticipate that any future native \nrpy-based codes will have even faster integration times. Key future technical developments will focus on integrating this distributed-memory capability with node-level GPU acceleration via the ongoing \nrpycuda project~\cite{Tootle:2025ikk} and extending code generation to support the multi-patch, multi-coordinate grids vital for tackling long inspirals and complex accretion physics.

Ultimately, by providing pathways for both efficient single-node execution (suitable for efforts like the \BHaH volunteer computing project) and massively parallel scalability on HPC clusters, the \nrpy framework equipped with the \superb extension offers a unified, open-source, and extensible platform. It empowers next-generation gravitational-wave source modeling by effectively leveraging computational resources ranging from desktops to leadership-class supercomputers, tailored to the demands of the scientific application.

\ack
\label{sec:acks}
This work was primarily supported by the NASA ATP award 80NSSC22K1898. ZBE also gratefully acknowledges support from NSF awards OAC-2004311, OAC-2411068, PHY-2108072, and PHY-2409654. TPJ gratefully acknowledges support from NASA FINESST-80NSSC23K1437. This research made use of Idaho National Laboratory’s High Performance Computing systems located at the Collaborative Computing Center and supported by the Office of Nuclear Energy of the U.S. Department of Energy and the Nuclear Science User Facilities under Contract No. DE-AC07-05ID14517. We also thank the \charmpp developers for their support and advice.  \charmpp/Converse was developed by the Parallel Programming Laboratory in the Department of Computer Science at the University of Illinois at Urbana-Champaign.

\section*{References}
\bibliographystyle{iopart-num}
\bibliography{ref}

\providecommand{\newblock}{}
\begin{thebibliography}{10}
\expandafter\ifx\csname url\endcsname\relax
  \def\url#1{{\tt #1}}\fi
\expandafter\ifx\csname urlprefix\endcsname\relax\def\urlprefix{URL }\fi
\providecommand{\eprint}[2][]{\url{#2}}

\bibitem{LIGOScientific:2018mvr}
Abbott B~P {\em et~al.\/} (LIGO Scientific, Virgo) 2019 {\em Phys. Rev. X\/} {\bf 9} 031040 (\textit{Preprint} \eprint{1811.12907})

\bibitem{LIGOScientific:2020ibl}
Abbott R {\em et~al.\/} (LIGO Scientific, Virgo) 2021 {\em Phys. Rev. X\/} {\bf 11} 021053 (\textit{Preprint} \eprint{2010.14527})

\bibitem{LIGOScientific:2021usb}
Abbott R {\em et~al.\/} (LIGO Scientific, VIRGO) 2024 {\em Phys. Rev. D\/} {\bf 109} 022001 (\textit{Preprint} \eprint{2108.01045})

\bibitem{KAGRA:2021vkt}
Abbott R {\em et~al.\/} (KAGRA, VIRGO, LIGO Scientific) 2023 {\em Phys. Rev. X\/} {\bf 13} 041039 (\textit{Preprint} \eprint{2111.03606})

\bibitem{Nitz:2021uxj}
Nitz A~H, Capano C~D, Kumar S, Wang Y~F, Kastha S, Sch\"afer M, Dhurkunde R and Cabero M 2021 {\em Astrophys. J.\/} {\bf 922} 76 (\textit{Preprint} \eprint{2105.09151})

\bibitem{Baiotti:2016qnr}
Baiotti L and Rezzolla L 2017 {\em Rept. Prog. Phys.\/} {\bf 80} 096901 (\textit{Preprint} \eprint{1607.03540})

\bibitem{Shibata:2019wef}
Shibata M and Hotokezaka K 2019 {\em Ann. Rev. Nucl. Part. Sci.\/} {\bf 69} 41--64 (\textit{Preprint} \eprint{1908.02350})

\bibitem{Pan:2009wj}
Pan Y, Buonanno A, Buchman L~T, Chu T, Kidder L~E, Pfeiffer H~P and Scheel M~A 2010 {\em Phys. Rev. D\/} {\bf 81} 084041 (\textit{Preprint} \eprint{0912.3466})

\bibitem{Ramos-Buades:2023ehm}
Ramos-Buades A, Buonanno A, Estell\'es H, Khalil M, Mihaylov D~P, Ossokine S, Pompili L and Shiferaw M 2023 {\em Phys. Rev. D\/} {\bf 108} 124037 (\textit{Preprint} \eprint{2303.18046})

\bibitem{Khan:2015jqa}
Khan S, Husa S, Hannam M, Ohme F, P\"urrer M, Jim\'enez~Forteza X and Boh\'e A 2016 {\em Phys. Rev. D\/} {\bf 93} 044007 (\textit{Preprint} \eprint{1508.07253})

\bibitem{Field5}
{Blackman} J, {Field} S~E, {Galley} C~R, {Szil{\'a}gyi} B, {Scheel} M~A, {Tiglio} M and {Hemberger} D~A 2015 {\em Phys. Rev. Lett.\/} {\bf 115} 121102 (\textit{Preprint} \eprint{1502.07758})

\bibitem{Varma:2019csw}
Varma V, Field S~E, Scheel M~A, Blackman J, Gerosa D, Stein L~C, Kidder L~E and Pfeiffer H~P 2019 {\em Phys. Rev. Research.\/} {\bf 1} 033015 (\textit{Preprint} \eprint{1905.09300})

\bibitem{Albanesi:2025txj}
Albanesi S, Gamba R, Bernuzzi S, Fontbuté J, Gonzalez A and Nagar A 2025 {\em arXiv preprint\/} ArXiv:2503.14580v1 (\textit{Preprint} \eprint{2503.14580})

\bibitem{Punturo2010}
{Punturo et al} 2010 {\em Classical and Quantum Gravity\/} {\bf 27} 194002

\bibitem{Reitze:2019dyk}
Reitze D {\em et~al.\/} 2019 {\em Bull. Am. Astron. Soc.\/} {\bf 51} 141 (\textit{Preprint} \eprint{1903.04615})

\bibitem{Evans2021}
Evans M {\em et~al.\/} 2021 A horizon study for cosmic explorer: Science, observatories, and community Technical Report CE-P2100003-v6 Cosmic Explorer Project (\textit{Preprint} \eprint{2109.09882}) \urlprefix\url{https://dcc.cosmicexplorer.org/CE-P2100003/public}

\bibitem{TianQin2015}
Luo J {\em et~al.\/} (TianQin) 2016 {\em Class. Quant. Grav.\/} {\bf 33} 035010 (\textit{Preprint} \eprint{1512.02076})

\bibitem{AmaroSeoane2017}
{P Amaro-Seoane et al} 2017 {\em arXiv e-prints\/} arXiv:1702.00786 (\textit{Preprint} \eprint{1702.00786})

\bibitem{Babak2021}
Babak S, Petiteau A and Hewitson M 2021 Lisa sensitivity and snr calculations Technical Note LISA-LCST-SGS-TN-001 LISA Consortium (\textit{Preprint} \eprint{2108.01167})

\bibitem{Purrer2020}
{P{\"u}rrer} M and {Haster} C~J 2020 {\em Phys. Rev. Research\/} {\bf 2} 023151 (\textit{Preprint} \eprint{1912.10055})

\bibitem{Ferguson2021}
{Ferguson} D, {Jani} K, {Laguna} P and {Shoemaker} D 2021 {\em Phys. Rev. D\/} {\bf 104} 044037 (\textit{Preprint} \eprint{2006.04272})

\bibitem{Jan2024}
Jan A, Ferguson D, Lange J, Shoemaker D and Zimmerman A 2024 {\em Phys. Rev. D\/} {\bf 110}(2) 024023 \urlprefix\url{https://link.aps.org/doi/10.1103/PhysRevD.110.024023}

\bibitem{AmaroSeoane2018}
{Amaro-Seoane} P 2018 {\em Phys. Rev. D\/} {\bf 98} 063018 (\textit{Preprint} \eprint{1807.03824})

\bibitem{Foucart2022}
{Foucart} F, {Laguna} P, {Lovelace} G, {Radice} D and {Witek} H 2022 {\em arXiv e-prints\/} arXiv:2203.08139 (\textit{Preprint} \eprint{2203.08139})

\bibitem{Schnetter2004}
{Schnetter} E, {Hawley} S~H and {Hawke} I 2004 {\em Classical and Quantum Gravity\/} {\bf 21} 1465--1488 (\textit{Preprint} \eprint{gr-qc/0310042})

\bibitem{Loffler2012}
Löffler F, Faber J, Bentivegna E, Bode T, Diener P, Haas R, Hinder I, Mundim B~C, Ott C~D, Schnetter E, Allen G, Campanelli M and Laguna P 2012 {\em Classical and Quantum Gravity\/} {\bf 29} 115001 ISSN 1361-6382 \urlprefix\url{http://dx.doi.org/10.1088/0264-9381/29/11/115001}

\bibitem{EinsteinToolkit2024_11}
Haas R {\em et~al.\/} 2024 {The Einstein Toolkit: The "Annie Jump Cannon" release (ET\_2024\_11)} released on December 1, 2024 \urlprefix\url{https://doi.org/10.5281/zenodo.14193969}

\bibitem{Pollney2011}
{Pollney} D, {Reisswig} C, {Schnetter} E, {Dorband} N and {Diener} P 2011 {\em Phys. Rev. D\/} {\bf 83} 044045 (\textit{Preprint} \eprint{0910.3803})

\bibitem{Brown2009_2}
{Brown} D, {Diener} P, {Sarbach} O, {Schnetter} E and {Tiglio} M 2009 {\em Phys. Rev. D\/} {\bf 79} 044023 (\textit{Preprint} \eprint{0809.3533})

\bibitem{Sperhake2007}
{Sperhake} U 2007 {\em Phys. Rev. D\/} {\bf 76} 104015 (\textit{Preprint} \eprint{gr-qc/0606079})

\bibitem{Brugmann2008}
{Br{\"u}gmann} B, {Gonz{\'a}lez} J~A, {Hannam} M, {Husa} S, {Sperhake} U and {Tichy} W 2008 {\em Phys. Rev. D\/} {\bf 77} 024027 (\textit{Preprint} \eprint{gr-qc/0610128})

\bibitem{Cao2008}
Cao Z, Yo H~J and Yu J~P 2008 {\em Phys. Rev. D\/} {\bf 78}(12) 124011 \urlprefix\url{https://link.aps.org/doi/10.1103/PhysRevD.78.124011}

\bibitem{Clough2015}
{Clough} K, {Figueras} P, {Finkel} H, {Kunesch} M, {Lim} E~A and {Tunyasuvunakool} S 2015 {\em Classical and Quantum Gravity\/} {\bf 32} 245011 (\textit{Preprint} \eprint{1503.03436})

\bibitem{Fernando2019}
{Fernando} M, {Neilsen} D, {Lim} H, {Hirschmann} E and {Sundar} H 2019 {\em SIAM Journal on Scientific Computing\/} {\bf 41} C97--C138 (\textit{Preprint} \eprint{1807.06128})

\bibitem{Fernando2022}
{Fernando} M, {Neilsen} D, {Zlochower} Y, {Hirschmann} E~W and {Sundar} H 2022 {\em arXiv e-prints\/} arXiv:2211.11575 (\textit{Preprint} \eprint{2211.11575})

\bibitem{Daszuta2021}
{Daszuta} B, {Zappa} F, {Cook} W, {Radice} D, {Bernuzzi} S and {Morozova} V 2021 {\em Astrophys. J. Suppl.\/} {\bf 257} 25 (\textit{Preprint} \eprint{2101.08289})

\bibitem{Daszuta2024}
{Daszuta} B and {Cook} W 2024 {\em arXiv e-prints\/} arXiv:2406.05126 (\textit{Preprint} \eprint{2406.05126})

\bibitem{Cook2025}
{Cook} W, {Daszuta} B, {Fields} J, {Hammond} P, {Albanesi} S, {Zappa} F, {Bernuzzi} S and {Radice} D 2025 {\em Astrophys. J. Suppl.\/} {\bf 277} 3 (\textit{Preprint} \eprint{2311.04989})

\bibitem{Szilagyi2009}
{Szil{\'a}gyi} B, {Lindblom} L and {Scheel} M~A 2009 {\em Phys. Rev. D\/} {\bf 80} 124010 (\textit{Preprint} \eprint{0909.3557})

\bibitem{Hilditch2016}
{Hilditch} D, {Weyhausen} A and {Br{\"u}gmann} B 2016 {\em Phys. Rev. D\/} {\bf 93} 063006 (\textit{Preprint} \eprint{1504.04732})

\bibitem{Bugner2015}
{Bugner} M, {Dietrich} T, {Bernuzzi} S, {Weyhausen} A and {Bruegmann} B 2015 {\em arXiv e-prints\/} arXiv:1508.07147 (\textit{Preprint} \eprint{1508.07147})

\bibitem{Kidder2016}
{Kidder} L~E, {Field} S~E, {Foucart} F, {Schnetter} E, {Teukolsky} S~A, {Bohn} A, {Deppe} N, {Diener} P, {H{\'e}bert} F, {Lippuner} J, {Miller} J, {Ott} C~D, {Scheel} M~A and {Vincent} T 2016 {\em arXiv e-prints\/} arXiv:1609.00098 (\textit{Preprint} \eprint{1609.00098})

\bibitem{Spectre}
Deppe N, Throwe W, Kidder L~E, Vu N~L, Nelli K~C, Armaza C, Bonilla M~S, H\'ebert F, Kim Y, Kumar P, Lovelace G, Macedo A, Moxon J, O'Shea E, Pfeiffer H~P, Scheel M~A, Teukolsky S~A, Wittek N~A {\em et~al.\/} 2024 \texttt{SpECTRE v2024.08.03} \href{https://doi.org/10.5281/zenodo.13207712}{10.5281/zenodo.13207712} \urlprefix\url{https://spectre-code.org}

\bibitem{Lovelace2025}
{Lovelace} G, {Nelli} K~C, {Deppe} N, {Vu} N~L, {Throwe} W, {Bonilla} M~S, {Carpenter} A, {Kidder} L~E, {Macedo} A, {Scheel} M~A, {Afram} A, {Boyle} M, {Ceja} A, {Giesler} M, {Habib} S, {Jones} K~Z, {Kumar} P, {Lara} G, {Melchor} D, {Mendes} I~B, {Mitman} K, {Morales} M, {Moxon} J, {O'Shea} E, {Pannone} K, {Pfeiffer} H~P, {Ramirez-Aguilar} T, {Sanchez} J, {Tellez} D, {Teukolsky} S~A and {Wittek} N~A 2025 {\em Classical and Quantum Gravity\/} {\bf 42} 035001 (\textit{Preprint} \eprint{2410.00265})

\bibitem{Tichy2023}
{Tichy} W, {Ji} L, {Adhikari} A, {Rashti} A and {Pirog} M 2023 {\em Classical and Quantum Gravity\/} {\bf 40} 025004 (\textit{Preprint} \eprint{2212.06340})

\bibitem{Hengrui2024}
{Zhu} H, {Fields} J, {Zappa} F, {Radice} D, {Stone} J, {Rashti} A, {Cook} W, {Bernuzzi} S and {Daszuta} B 2024 {\em arXiv e-prints\/} arXiv:2409.10383 (\textit{Preprint} \eprint{2409.10383})

\bibitem{Husa2006}
{Husa} S, {Hinder} I and {Lechner} C 2006 {\em Computer Physics Communications\/} {\bf 174} 983--1004 (\textit{Preprint} \eprint{gr-qc/0404023})

\bibitem{Palenzuela2021}
{Palenzuela} C, {Mi{\~n}ano} B, {Arbona} A, {Bona-Casas} C, {Bona} C and {Mass{\'o}} J 2021 {\em Computer Physics Communications\/} {\bf 259} 107675 (\textit{Preprint} \eprint{2010.00902})

\bibitem{Ruchlin2018}
{Ruchlin} I, {Etienne} Z~B and {Baumgarte} T~W 2018 {\em Phys. Rev. D\/} {\bf 97} 064036 (\textit{Preprint} \eprint{1712.07658})

\bibitem{Pretorius:2005gq}
Pretorius F 2005 {\em Phys. Rev. Lett.\/} {\bf 95} 121101 (\textit{Preprint} \eprint{gr-qc/0507014})

\bibitem{Campanelli:2005dd}
Campanelli M, Lousto C~O, Marronetti P and Zlochower Y 2006 {\em Phys. Rev. Lett.\/} {\bf 96} 111101 (\textit{Preprint} \eprint{gr-qc/0511048})

\bibitem{Baker:2005vv}
Baker J~G, Centrella J, Choi D~I, Koppitz M and van Meter J 2006 {\em Phys. Rev. Lett.\/} {\bf 96} 111102 (\textit{Preprint} \eprint{gr-qc/0511103})

\bibitem{GarfinkleGeneralizedHarmonicpre}
{Garfinkle} D 2002 {\em Phys. Rev. D\/} {\bf 65} 044029 (\textit{Preprint} \eprint{gr-qc/0110013})

\bibitem{Pretorius:2004jg}
Pretorius F 2005 {\em Class. Quant. Grav.\/} {\bf 22} 425--452 (\textit{Preprint} \eprint{gr-qc/0407110})

\bibitem{Shibata1995}
Shibata M and Nakamura T 1995 {\em Phys. Rev. D\/} {\bf 52}(10) 5428--5444 \urlprefix\url{https://link.aps.org/doi/10.1103/PhysRevD.52.5428}

\bibitem{Baumgarte1998}
{Baumgarte} T~W and {Shapiro} S~L 1998 {\em Phys. Rev. D\/} {\bf 59} 024007 (\textit{Preprint} \eprint{gr-qc/9810065})

\bibitem{Brown2009}
{Brown} J~D 2009 {\em Phys. Rev. D\/} {\bf 79} 104029 (\textit{Preprint} \eprint{0902.3652})

\bibitem{Zlochower2012}
{Zlochower} Y, {Ponce} M and {Lousto} C~O 2012 {\em Phys. Rev. D\/} {\bf 86} 104056 (\textit{Preprint} \eprint{1208.5494})

\bibitem{Etienne2014}
{Etienne} Z~B, {Baker} J~G, {Paschalidis} V, {Kelly} B~J and {Shapiro} S~L 2014 {\em Phys. Rev. D\/} {\bf 90} 064032 (\textit{Preprint} \eprint{1404.6523})

\bibitem{Black:2025orbit}
{Black} W~K, {Neilsen} D, {Hirschmann} E~W, {Van Komen} D~F and {Fernando} M 2025 {\em arXiv e-prints\/} arXiv:2502.20282 (\textit{Preprint} \eprint{2502.20282})

\bibitem{Baumgarte2013}
{Baumgarte} T~W, {Montero} P~J, {Cordero-Carri{\'o}n} I and {M{\"u}ller} E 2013 {\em Phys. Rev. D\/} {\bf 87} 044026 (\textit{Preprint} \eprint{1211.6632})

\bibitem{nrpy_web}
 2025 {\texttt{NRPy} GitHub repository} \url{https://github.com/nrpy/nrpy}

\bibitem{sympy}
Meurer A {\em et~al.\/} 2017 {\em PeerJ Computer Science\/} {\bf 3} e103:1--27 ISSN 2376-5992 \urlprefix\url{https://doi.org/10.7717/peerj-cs.103}

\bibitem{Etienne2024}
{Etienne} Z~B 2024 {\em Phys. Rev. D\/} {\bf 110} 064045 (\textit{Preprint} \eprint{2404.01137})

\bibitem{Montero2012}
{Montero} P~J and {Cordero-Carri{\'o}n} I 2012 {\em Phys. Rev. D\/} {\bf 85} 124037 (\textit{Preprint} \eprint{1204.5377})

\bibitem{kale1993charmpp}
Kale L~V and Krishnan S 1993 {Charm++}: A portable concurrent object oriented system based on {C++} {\em Proceedings of the 8th Annual Conference on Object-Oriented Programming Systems, Languages, and Applications (OOPSLA)\/} (Washington, D.C., USA: Association for Computing Machinery) pp 91--108 \urlprefix\url{https://doi.org/10.1145/165854.165873}

\bibitem{charmpp}
{The Charm++ Development Team} 2025 {Charm++: The Parallel Programming System} \url{https://github.com/UIUC-PPL/charm} accessed: 2025-04-13

\bibitem{Mewes:2018szi}
Mewes V, Zlochower Y, Campanelli M, Ruchlin I, Etienne Z~B and Baumgarte T~W 2018 {\em Phys. Rev. D\/} {\bf 97} 084059 (\textit{Preprint} \eprint{1802.09625})

\bibitem{Mewes:2020vic}
Mewes V, Zlochower Y, Campanelli M, Baumgarte T~W, Etienne Z~B, Lopez~Armengol F~G and Cipolletta F 2020 {\em Phys. Rev. D\/} {\bf 101} 104007 (\textit{Preprint} \eprint{2002.06225})

\bibitem{Ji:2023tok}
Ji L, Mewes V, Zlochower Y, Ennoggi L, Armengol F~G~L, Campanelli M, Cipolletta F and Etienne Z~B 2023 {\em Phys. Rev. D\/} {\bf 108} 104005 (\textit{Preprint} \eprint{2305.01537})

\bibitem{Alic:2011gg}
Alic D, Bona-Casas C, Bona C, Rezzolla L and Palenzuela C 2012 {\em Phys. Rev. D\/} {\bf 85} 064040 (\textit{Preprint} \eprint{1106.2254})

\bibitem{Mosta:2013gwu}
M\"osta P, Mundim B~C, Faber J~A, Haas R, Noble S~C, Bode T, L\"offler F, Ott C~D, Reisswig C and Schnetter E 2014 {\em Class. Quant. Grav.\/} {\bf 31} 015005 (\textit{Preprint} \eprint{1304.5544})

\bibitem{senr_web}
 2019 Deprecated {N}{R}{P}y+ repository, hosted on bitbucket \url{https://bitbucket.org/zach_etienne/nrpy}

\bibitem{Schnetter:2003rb}
Schnetter E, Hawley S~H and Hawke I 2004 {\em Class. Quant. Grav.\/} {\bf 21} 1465--1488 (\textit{Preprint} \eprint{gr-qc/0310042})

\bibitem{CarpetCode:web}
 2024 Carpet: Adaptive mesh refinement for the {Cactus Framework} \url{http://www.carpetcode.org/}

\bibitem{Jacques:2024pxh}
Jacques T~P, Cupp S, Werneck L~R, Tootle S~D, Hamilton M~C~B and Etienne Z~B 2024 {\em arXiv preprint\/} (\textit{Preprint} \eprint{2412.03659})

\bibitem{Tootle:2025ikk}
Tootle S~D, Werneck L~R, Assumpcao T, Jacques T~P and Etienne Z~B 2025 {\em arXiv preprint\/} (\textit{Preprint} \eprint{2501.14030})

\bibitem{Tichy2007}
{Tichy} W and {Marronetti} P 2007 {\em Phys. Rev. D\/} {\bf 76} 061502 (\textit{Preprint} \eprint{gr-qc/0703075})

\bibitem{Marronetti2008}
{Marronetti} P, {Tichy} W, {Br{\"u}gmann} B, {Gonz{\'a}lez} J and {Sperhake} U 2008 {\em Phys. Rev. D\/} {\bf 77} 064010 (\textit{Preprint} \eprint{0709.2160})

\bibitem{Bona1995}
{Bona} C, {Mass{\'o}} J, {Seidel} E and {Stela} J 1995 {\em Phys. Rev. Lett.\/} {\bf 75} 600--603 (\textit{Preprint} \eprint{gr-qc/9412071})

\bibitem{Alcubierre2003}
{Alcubierre} M, {Br{\"u}gmann} B, {Diener} P, {Koppitz} M, {Pollney} D, {Seidel} E and {Takahashi} R 2003 {\em Phys. Rev. D\/} {\bf 67} 084023 (\textit{Preprint} \eprint{gr-qc/0206072})

\bibitem{Schnetter2010}
{Schnetter} E 2010 {\em Classical and Quantum Gravity\/} {\bf 27} 167001 (\textit{Preprint} \eprint{1003.0859})

\bibitem{Assumpcao2022}
{Assump{\c{c}}{\~a}o} T, {Werneck} L~R, {Pierre Jacques} T and {Etienne} Z~B 2022 {\em Phys. Rev. D\/} {\bf 105} 104037 (\textit{Preprint} \eprint{2111.02424})

\bibitem{GottliebShuTadmorSSPRK}
Gottlieb S, Shu C~W and Tadmor E 2001 {\em SIAM Review\/} {\bf 43} 89--112 \urlprefix\url{https://www.jstor.org/stable/3649684}

\bibitem{Ansorg2004}
{Ansorg} M, {Br{\"u}gmann} B and {Tichy} W 2004 {\em Phys. Rev. D\/} {\bf 70} 064011 (\textit{Preprint} \eprint{gr-qc/0404056})

\bibitem{Alcubierre2003_2}
{Alcubierre} M 2003 {\em Classical and Quantum Gravity\/} {\bf 20} 607--623 (\textit{Preprint} \eprint{gr-qc/0210050})

\bibitem{Alcubierre2005}
{Alcubierre} M 2005 {\em Classical and Quantum Gravity\/} {\bf 22} 4071--4081 (\textit{Preprint} \eprint{gr-qc/0503030})

\bibitem{Stein:2019mop}
Stein L~C 2019 {\em J. Open Source Softw.\/} {\bf 4} 1683 (\textit{Preprint} \eprint{1908.10377})

\bibitem{Anninos1993}
Anninos P, Hobill D, Seidel E, Smarr L and Suen W~M 1993 {\em Phys. Rev. Lett.\/} {\bf 71}(18) 2851--2854 \urlprefix\url{https://link.aps.org/doi/10.1103/PhysRevLett.71.2851}

\bibitem{Kale2006}
Kalé L~V, Zheng G, Lee C~W and Kumar S 2006 {\em Future Generation Computer Systems\/} {\bf 22} 347--358 ISSN 0167-739X \urlprefix\url{https://www.sciencedirect.com/science/article/pii/S0167739X04002262}

\end{thebibliography}

\end{document}